\documentclass[pdftex,twocolumn,epjc3]{svjour3} 
\RequirePackage[T1]{fontenc}
\smartqed
\RequirePackage{graphicx}
\RequirePackage{mathptmx}
\RequirePackage{flushend}
\RequirePackage[numbers,sort&compress]{natbib}
\RequirePackage{lipsum,amsmath,multicol}
\RequirePackage[colorlinks,citecolor=blue,urlcolor=blue,linkcolor=blue,draft]{hyperref}
\RequirePackage{amsmath}
\RequirePackage{epstopdf}
\RequirePackage{mathtools, cuted}
\RequirePackage{float}

\journalname{Eur. Phys. J. C}

\usepackage{etoolbox}
\begin{document}

\title{Bounds on spinning particles in their innermost stable circular orbits around rotating braneworld black hole
}


\author{Ulises Nucamendi\thanksref{e1,addr1,addr2}
        \and
        Ricardo Becerril\thanksref{e2,addr1} 
        \and
        Pankaj Sheoran\thanksref{e3,addr1}
}

\thankstext{e1}{e-mail: unucamendi@gmail.com}
\thankstext{e2}{e-mail: ricardo.becerril@umich.mx}
\thankstext{e3}{e-mail: pankaj.sheoran@umich.mx}

\institute{Instituto de F\'{\i}sica y Matem\'{a}ticas, Universidad Michoacana de San Nicol\'{a}s de Hidalgo,
Edificio C-3, 58040 Morelia, Michoac\'{a}n, M\'{e}xico.\label{addr1}
          \and
          Mesoamerican Centre for Theoretical Physics, Universidad Aut\'onoma de Chiapas.
Ciudad Universitaria, Carretera Zapata Km. 4, Real del Bosque (Ter\'{a}n), 29040, Tuxtla Guti\'{e}rrez, Chiapas, M\'{e}xico.
\label{addr2}
}

\date{Received: date / Accepted: date}

\maketitle

\begin{abstract}
We study the innermost stable circular orbit (ISCO) of a spinning test particle moving in the vicinity of an axially symmetric rotating braneworld black hole (BH). We start with the description of the event horizon, static limit surface and ergosphere region of such BH and bring out the effect of tidal charge parameter on ergosphere. It is found that the ISCO of rotating braneworld BH is very sensitive to braneworld BH parameter $\mathit{C}$ (also known as tidal charge parameter) in addition to its rotation parameter. We further discovered that the orbital radius of the spinning test particles changes non-monotonously with the braneworld BH tidal charge parameter. It is found that for rotating brane-world BH the allowed range of the particle spin grows as the tidal charge parameter $\mathit{C}$ decreases, in contrast with the Kerr-Newman BH. We also found the similar behavior of the particle's spin for the braneworld Reissner-Nordstr$\ddot{o}$m ($\mathit{C}<0$) BH in contrast with its counterpart having ($\mathit{C}>0$).
\end{abstract}


\maketitle

\section{Introduction}\label{scheme1}

In this era of advanced laser interferometers (i.e., Laser Interferometer Gravitational Wave Observatory (LIGO) \cite{Harry:2010zz}, Virgo \cite{Accadia:2011zzc}, and KAGRA \cite{Somiya:2011np}) and high resolution telescopes (i.e. Event Horizon Telescope (EHT)) \cite{Akiyama:2019cqa,Akiyama:2019brx,Akiyama:2019sww,Akiyama:2019bqs,Akiyama:2019fyp,Akiyama:2019eap}, the dynamics of bodies (point-like or extended) around a central black hole (BH) is of indispensable interest because a compact body such as a BH or neutron star of few solar mass orbiting around a central massive BH (of mass $\sim10^4$ to $10^6$ $M_{\odot}$) is usually referred to an extreme mass ratio inspiral and is a propitious source of gravitational waves for the space-based interferometer eLISA \cite{Seoane:2013qna} and DECIGO \cite{Kawamura:2006up} in the near future. In these scenarios of extreme mass ratios, the numerical relativity approach is not very efficient \cite{Cardoso:2014uka} and the approximation methods like the effective one body formalism \cite{Buonanno:1998gg,Damour:2008yg} become handy.

Previous to these theoretical advancements, the dynamics of a non-spinning particle moving on a geodesic around a Schwarzschild BH was first studied by Kaplan in \cite{Kaplan}.
Since then the dynamics of non-spinning particles has been studied vastly by researchers
\cite{Shahrear:2007zz,Hackmann:2010zz,Pugliese:2011xn,Pugliese:2010ps,Pugliese:2011py,Pugliese:2010he,Slany:2013ora,Hernandez-Pastora:2013pia,Wunsch:2013st,Hod:2013vi,Chakraborty:2013kza,Hod:2014tpa,Zaslavskii:2014mqa,Lee:2017fbq,Chartas:2016ckd,Chaverri-Miranda:2017gxq,Sharif:2018cac,Chakraborty:2019rna}. The study of non-spinning particles in the vicinity of different BHs showed that the motion of the massive or massless test particles gets affected by the BHs parameters like mass, charge and rotation. In addition to these parameters, it is also found that if the central object (i.e. BH) is inspired by alternative theories of gravity then the motion of the non-spinning particles in the vicinity of BH gets influenced by the extra parameter known as the deviation parameter \cite{Lee:2017fbq,Chaverri-Miranda:2017gxq,Chakraborty:2019rna,Sharif:2018cac}.
On the hand, the dynamics of spinning particles around the BHs has not received as much attention as it needs to be.

The study of spinning particles started with the pioneering works of Mathisson \cite{Mathisson}, Papapetrou \cite{Papapetrou1951a,Papapetrou1951c} and Dixon \cite{Dixon} (MPD), where they developed the dynamical equations of the spinning particles moving in curved backgrounds by considering the ``pole-dipole" approximation only. The theory of the spinning particles was further developed by several authors \cite{Hanson1974,phdthesis,Tod:1976ud,Hojman1977,Chicone:2005jj,Mashhoon:2006fj,Hojman:2012me}, where it was showed that the four-velocity vector and the corresponding conjugate momentum vector are not parallel to each other which is in contrast with the case of the nonspinning particle where these two corresponding quantities are parallel. In \cite{phdthesis,Hojman:2012me} it was also showed that, for the case of the spinning particle, the four velocity might change from timelike to spacelike if the spin of the particle is greater than a certain critical limit whereas the conjugate momentum always remains timelike along its trajectory and satisfying the conservation of mass relation ($P_{\mu}P^{\mu}=-m^2$). It is relevant to point out here that the study of the dynamics of the spinning particle in a curved spacetime by considering the ``pole-dipole" approximation is valid only for homogeneous fields and fails when nonhomogeneous fields are taken into consideration. However, the study of spinning particles in curved spacetimes is extended to higher order-corrections in \cite{Steinhoff:2009tk,Steinhoff:2012rw}. It was showed in \cite{Deriglazov:2015zta,Deriglazov:2015wde,Ramirez:2017pmp,Deriglazov:2017jub} that the four velocity of a spinning particle will always remain timelike and never transforms to spacelike (i.e. it avoids the Superluminal problem) if one considers the coupling between spin and gravity via the gravitomagnetic moment, that is to say, that the superluminal regime can be avoided if multipole effects are taken into account. Taking into consideration the effect of multipole moments becomes important when one wishes to describe the self-gravitating compact bodies and gravitational radiation emitted by them. Recently, in \cite{Han:2008zzf,Han:2010tp,Warburton:2017sxk} the authors have studied the gravitational radiation emitted by the spinning particles moving in curved backgrounds as well as the associated chaos effect.

From the literature \cite{phdthesis,Tod:1976ud,Hojman1977,Han:2008zzf,Han:2010tp,Warburton:2017sxk,Suzuki:1997by,Semerak:1999qc,Hojman:2012me,Chicone:2005jj,Mashhoon:2006fj,Kyrian:2007zz,Singh:2008qr,Obukhov:2010kn,Plyatsko:2013xza,Hackmann:2014tga,Zalaquett2014,Jefremov:2015gza,Harms:2016ctx,Lukes-Gerakopoulos:2017vkj,Zhang:2017nhl,Deriglazov:2015wde,Deriglazov:2015zta,Ramirez:2017pmp,Deriglazov:2017jub,Armaza2016}, it is well understood that a nonspinning particle will follow a geodesic in the vicinity of a BH. On the contrary, if the interactions (i.e. gravitational self-force) of a test particle are taken into account then its trajectory is no longer a geodesic \cite{Warburton:2011hp,Isoyama:2014mja,vandeMeent:2016hel}. Also, in the case of a spinning test particle, the motion is non geodesic due to an additional force known as the spin-curvature force \cite{Hanson1974,Wald:1972} which comes into play due to the interaction of spin of the particle and the curvature of spacetime around a massive central object. In this work we only consider the spin-curvature coupling and discard all other reaction of the particle with the background of a BH and study the ISCO of a spinning test particle in its vicinity. The ISCO of a spinning particle around a Schwarzchild and Kerr BH was first studied by Suzuki and Maeda in their pioneering work \cite{Suzuki:1997by}. Later on, the ISCO parameters of the spinning particles  were obtained for a slowly rotating Kerr BH up to quadratic order, in terms of rotation parameter $a$ and particle's spin by Jefremov et. al \cite{Jefremov:2015gza}.

The study of the inner most stable circular orbit (ISCO) is very important from the point of view of the gravitational wave astronomy because the circular orbits which are located inside the ISCO are unstable under a perturbation, and then they can be taken as an initial point for the final merger of any binary system \cite{Shibata:1998ih,Zdunik:2000qn,Baumgarte:2001ab,Grandclement:2001ed,Miller:2003vc,Marronetti:2003hx,Campanelli:2006gf}. Also, the study of ISCO for a given BH tells about the properties of its background geometry because the motion of a particle depends on its mass, charge, rotation and extra deviation parameters (coming from alternative theories of gravity).

Since the study of spinning particle dynamics has been done mostly in the context of Einstein's theory but scarcely for alternative theories of gravity,
in this work we study the ISCO of a spinning massive test particle with arbitrary spin $s$ in the vicinity of a rotating braneworld BH \cite{Dadhich:2000am,Aliev:2005bi,Schee:2008fc,Stuchlik:2008fy,Stuchlik:2017rir} which has an additional parameter $\mathit{C}$ known as the braneworld tidal charge parameter in addition to the usual mass $M$ and rotation $a$ parameters.

The braneworld models are an effective four-dimensional version of higher-dimensional string theory \cite{Randall:1999vf,Randall:1999ee}. According to these models, our physical universe nests on a 3-brane of a higher dimensional spacetime while gravity enters as an extra spatial dimension \cite{ArkaniHamed:1998rs}. Therefore, studying the behavior of gravity in braneworld models shades light on the physical signature of higher dimensions on our four-dimensional physical world. In this context,
the braneworld BHs are interesting to study. Additionally, the study of brane-world BHs is fascinating in many other ways, one that the non-rotating braneworld BHs can be characterized with a Reissner-Nordstr\"{o}m type geometry \cite{Dadhich:2000am}, and the axially symmetric rotating braneworld BHs as a Kerr-Newman type geometry \cite{Aliev:2005bi}. Second, the tidal charge parameter can have both positive and negative values \cite{Dadhich:2000am,Aliev:2005bi} unlike in Einstein-Maxwell theory where the square of the electric charge is always positive. These exciting properties of braneworld BHs gained a lot of attention from researchers in the past decade and were studied in works related to accretion phenomenon \cite{Stuchlik:2008fy}, solar system tests \cite{Boehmer:2008zh}, quasiperiodic oscillations (QPOs) \cite{Kotrlova:2008xs}, shadows of BH\cite{Amarilla:2011fx}, gravitational lensing \cite{BinNun:2009jr} and many more topics \cite{Schee:2008fc,Stuchlik:2017rir,Schee:2008kz,Aliev:2009cg}. More recently, in \cite{Vagnozzi:2019apd} the shadow of M87* was used to constrain the curvature radius of rotating braneworld BH and in particular about the tidal charge parameter $\mathit{C}$.

In this context and for simplicity, we consider only the ``pole-dipole" estimation and numerically investigate the ph-ysical behavior for the ISCO parameters (radius $r$, energy $E$ and orbital angular momentum component in the $z$-direction $L_{z}$) using the superluminal constraint condition and imposing the Tulczyjew spin-supplementary condition (TSSC) for the spinning massive test particle in the rotating braneworld BH background. To be consistent, we showed that our results for the ISCO parameters of non-extremal cases of Kerr and Kerr-Newman like BHs ($\mathit{C}>0$) match exactly with the results obtained in \cite{Jefremov:2015gza,Zhang:2017nhl} for Kerr and Kerr-Newman BHs.

We organized our paper as follows, in Sec. \ref{G_eq_spin} we start with a brief review of the MPD equations (equations of motion (EOM) of a spinning particle) in a curved background. Next, we divide the Sec. \ref{scheme1} into two subsections: In the first subsection, we study the behavior of the event horizon, static limit surface and ergosphere (Fig. \ref{fig1_ergo}) for different values of the tidal charge parameter $\mathit{C}$ and numerically present the bound on the rotation $a$ and tidal charge parameters $\mathit{C}$ (Fig. \ref{fig2_cont}) of rotating braneworld BH and in the second subsection, we find its conserved quantities and present the EOM for the spinning particles moving around it. In Sec. \ref{Effective_potential}, we find the expression for the effective potential and show its behavior in Fig. \ref{fig3_veff} for different values of the particle spin ${\bf S}$ and the parameter $\mathit{C}$. In Sec. \ref{ISCO}, we study numerically the behavior of ISCO parameters for spinning particles moving in the vicinity of rotating braneworld BH. It comes out that for rotating braneworld BH the range of the particle spin, for which the behavior of ISCO parameters are physical, increases as the parameter $\mathit{C}$ decreases. In this section we also present a summary of the results that we have obtained from Figs. \ref{fig4_ISCO_parameters_co_rotating}, \ref{fig5_ISCO_parameters_counter_rotating}, \ref{fig6_Phase_Plot1} and \ref{fig7_Phase_Plot2}. We summarize and give conclusions about our work in Sec. \ref{Conclusion}. Finally, in the \ref{appendix} we showed the explicit form of the equations that are used to study the behavior of ISCO parameters.

Throughout the paper, we have worked with (${-}$,+,+,+) signature and fixed the fundamental constants ($c$ and $G$) to unity. Additionally, the transformation we have used for projecting any four-vector in the tetrad frame is as follows: $x^{(a)}=e^{(a)}_{\nu}x^{\nu}$ where the indices in curve brackets indicate tetrad components while greek indices mean spacetime components.

\section{The General equations of a spinning test particle}{\label{G_eq_spin}}
In this section, we briefly review the general EOM of an extended body for a spinning test particle in a curved spacetime in the pole-dipole approximation. These EOMs were first obtained by Mathisson \cite{Mathisson} and Papapetrou \cite{Papapetrou1951a,Papapetrou1951c} and later fine tuned by Dixon \cite{Dixon}. The final form of the set of EOMs reads
\begin{eqnarray}
\frac{DP^{\mu}}{D\tau}&=&-\frac{1}{2}R^{\mu}_{\nu \rho \sigma}U^{\nu}S^{\rho \sigma}\label{dp},\\
\frac{DS^{\mu \nu}}{D\tau}&=& P^{\mu}U^{\nu}-U^{\mu}P^{\nu}=S^{\mu \sigma}u^{\nu}_{\sigma}-u^{\mu \sigma}S^{\nu}_{\sigma}\label{ds},
\end{eqnarray}
where, $\tau$, $U^{\mu}\equiv dx^{\mu}/d\tau$, $u^{\mu\nu}$, $P^{\mu}$, $S^{\mu\nu}$ and $R^{\mu}_{\nu \rho \sigma}$ represent the affine parameter of the orbit, four-velocity, angular velocity tensor, conjugate momentum vector, spin angular momentum tensor and Riemann tensor of a curved spacetime, respectively. The EOMs mentioned above were obtained by considering the curved spacetime lagrangian ($\mathcal{L}=\mathcal{L}(c_{1},c_{2},$ $c_{3},c_{4}$) with the ``pole-dipole" approximation. The quantities $c_{1},c_{2},c_{3}$ and $c_{4}$ are four independent invariants \cite{phdthesis,Hojman1977} defined as:
\begin{eqnarray}
c_1&=&U^\mu U_\mu,\nonumber\\
c_2&=&u^{\mu\nu}u_{\mu\nu}=-\mathrm{Tr}(u^2),\nonumber\\
c_3&=&U_\alpha u^{\alpha\beta}u_{\beta\gamma}U^\gamma,\nonumber\\
 c_4&=&g_{\mu\nu}g_{\rho\tau}g_{\alpha\beta}g_{\gamma\delta}u^{\delta\mu}u^{\nu\rho}
u^{\tau\alpha}u^{\beta\gamma}.
\end{eqnarray}
particles in curved spacetime do not follow geodesic motion due to the coupling between the spin tensor and the curvature of the background geometry and hence present non zero acceleration as shown in Eqn. (\ref{dp}). The case of spinning particles is considered as an open system because in this case the $P^{\mu}$ and $U^{\mu}$ are not proportional to each other which gives rise to fourteen unknown variables (i.e. four for $P^{\mu}$ and $U^{\mu}$ each and six for the spin tensor $S^{\mu \nu}$), while we have only ten equations at hand. Hence, to close this open system we need extra constraint conditions (also known as the spin supplementary condition (SSC)). In the literature available, there are three widely studied SSCs that can be used to close the open system of equations of motion, namely: ($i$) the Papapetrou and Corinaldesi SSC $S^{0 i}=0$ leading to the no dipolar mass moment condition, $(ii)$ The Mathisson-Pirani SSC $S^{\mu \nu}U_{\nu}=0$ implying that orbits are helical in nature, $(iii)$ the Tulczyjew-Dixon SSC $S^{\mu \nu}P_{\nu}=0$ which leads to an exact solution of Eqs. (\ref{dp}) and (\ref{ds}) which in turn gives conservation of the mass $M$ and the spin $S$ of the spinning particle as shown in \cite{Hojman:2012me} and defined by the relation:
\begin{eqnarray}
m^{2}&\equiv &-P^{\mu}P_{\mu},\label{m2}\\
S^{2}&\equiv &\frac{1}{2}S^{\mu \nu}S_{\mu \nu}.\label{s2}
\end{eqnarray}

In this work, we use the Tulczyjew-Dixon SSC. For convenience we also normalize the affine parameter $\tau$ with the help of a normalized momentum ($V^{\mu} \equiv P^{\mu}/m$) as
\begin{equation}\label{nor_v}
V^{\mu}U_{\mu}=-1
\end{equation}
Now, using the Eqs. (\ref{s2}) and (\ref{nor_v}) together with the Tulczyjew-Dixon SSC, one can obtain the following relation between $U^{\mu}$ and $V^{\mu}$ \cite{Hojman1977}
\begin{equation}\label{uv}
U^{\mu}-V^{\mu}=\frac{2S^{\mu \nu}V^{\sigma}R_{\nu \sigma \gamma \delta}S^{\gamma \delta}}{4m^{2}+R_{\alpha \beta \kappa \lambda}S^{\alpha \beta}S^{\kappa \lambda}}.
\end{equation}
It is easy to see from the above equation that the four-velocity and four-momentum are not parallel anymore due to existence of the particle spin. Thus, to study the dynamics of the spinning particle in the background of BH inspired by the modified theories of gravity we need to solve the Eqs. (\ref{dp}), (\ref{ds}), (\ref{uv}) and the Tulczyjew-Dixon SSC.

\section{The metric for the rotating braneworld black hole and conserved quantities of spinning particles orbits}{\label{scheme1}}
\subsection{The Metric}
The static BH solution for the braneworld scenario was first found by Dadhich et. al. in \cite{Dadhich:2000am}. The metric is obtained by solving the 5D Einstein field equations constrained to the  3D-brane. We start this section by introducing the metric for a rotating BH in the Randall-Sundrum braneworld scenario and localized to a 3D-brane which was found in \cite{Aliev:2005bi}, the metric can be written in terms of Boyer-Lindquist coordinates as
\begin{eqnarray}
\label{metric}
       ds^2 &=& -\left(\frac{-a^2\sin^{2}\theta+\Delta}{\Sigma}\right) dt^2
        +\frac{\Sigma}{\Delta}dr^{2}
        \nonumber\\
        &&
        -2\left(\frac{r^2+a^2-\Delta}{\Sigma}\right)a \sin^{2}\theta dt d\phi
      +\Sigma d\theta^2
      \nonumber\\
     &&+\left[\frac{-\Delta\; a^{2}\sin^2\theta +\left(r^2+a^2\right)^2}{\Sigma}\right]\sin^2\theta d\phi^2,
\end{eqnarray}
where
\begin{eqnarray}
 \Delta&=&r^2-2M r + a^2 + \mathit{C},\nonumber\\
 \Sigma&=&r^2+a^2\cos^2\theta.
\end{eqnarray}

Here $a$ and $\mathit{C}$ are the rotation and tidal charge parameters respectively. The parameter $\mathit{C}$ gives rise to three class of BH metrics:

(i) $\mathit{C}>0$ corresponding to the Kerr-Newman like metric of Einstein-Maxwell theory,
(ii) $\mathit{C}=0$ corresponding to the standard Kerr metric of GR and
(iii) $\mathit{C}<0$ corresponding to the Kerr-Newman like braneworld metric with negative tidal effects.

In complete analogy to the Kerr-Newman BH, the metric (\ref{metric}) possesses two major surfaces, namely the event horizon 
\begin{figure*}
\begin{tabular}{c c c c c}
\includegraphics[scale=0.66]{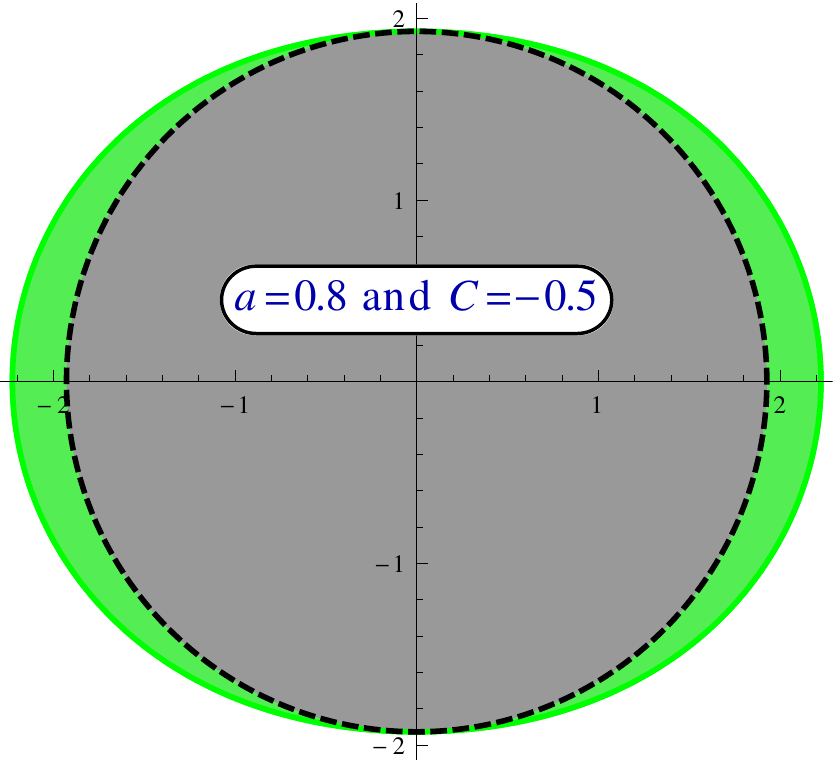}\hspace{0cm}
&\includegraphics[scale=0.66]{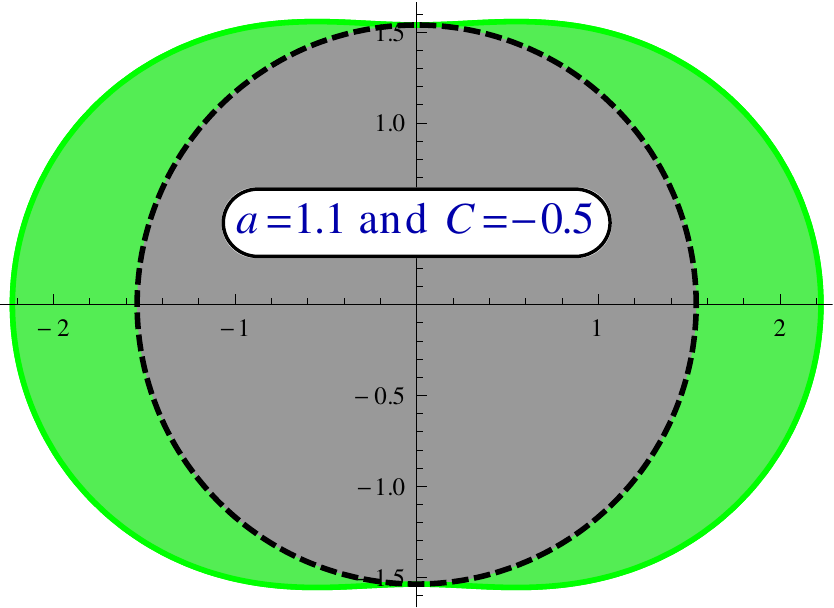}\hspace{0cm}
&\includegraphics[scale=0.66]{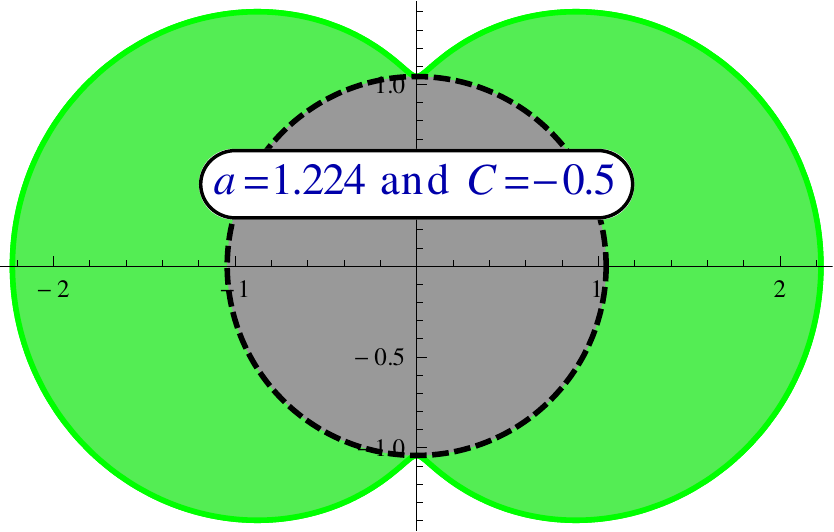}\hspace{0cm}
\\
\includegraphics[scale=0.66]{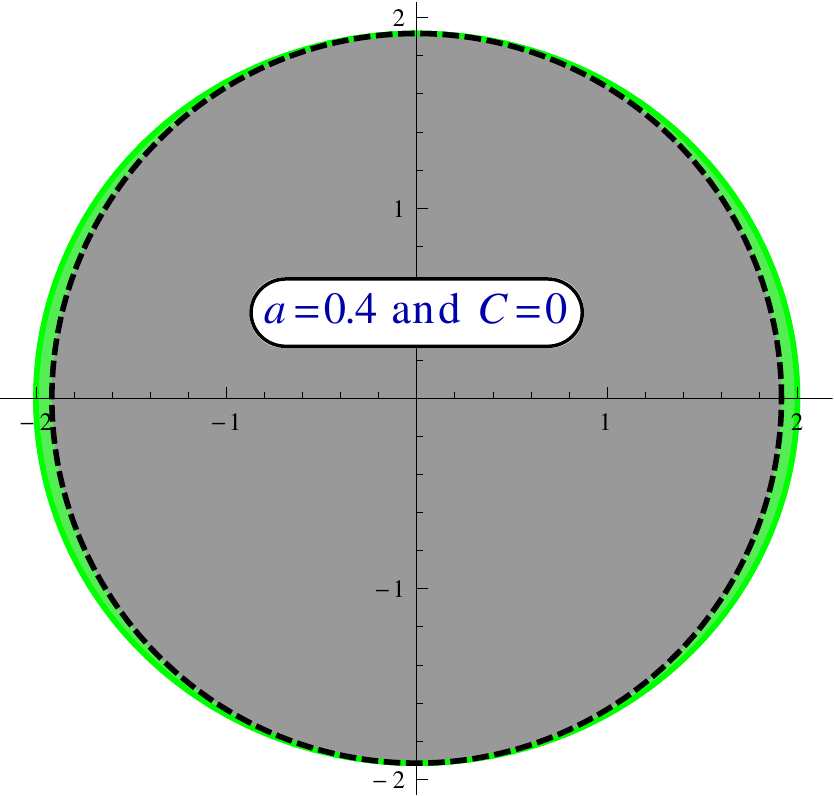}\hspace{0cm}
&\includegraphics[scale=0.66]{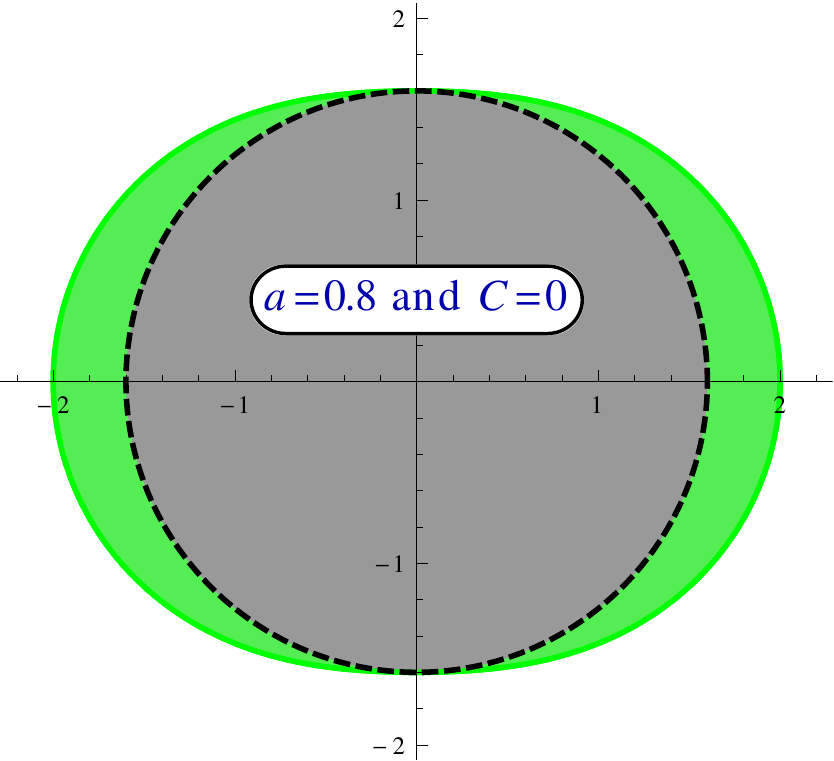}\hspace{0cm}
&\includegraphics[scale=0.66]{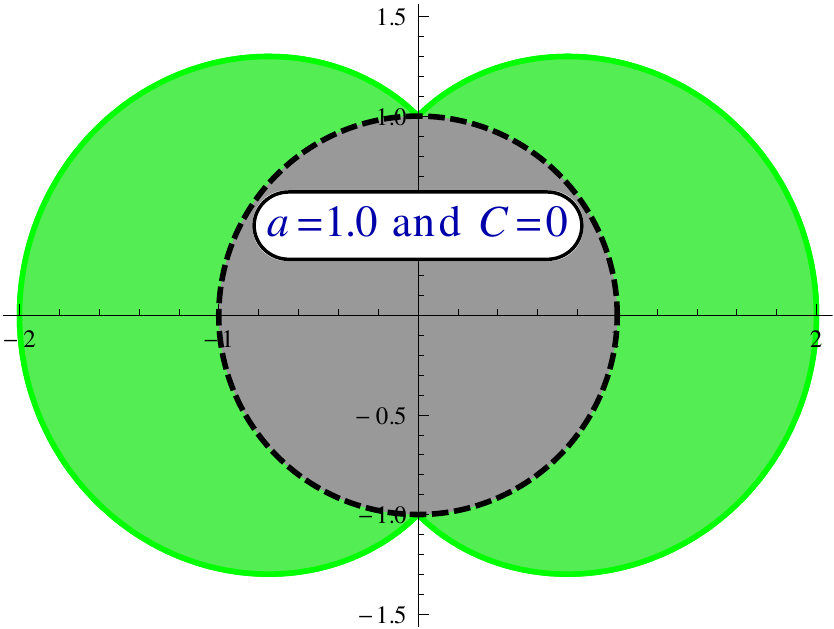}\hspace{0cm}
\\
\includegraphics[scale=0.66]{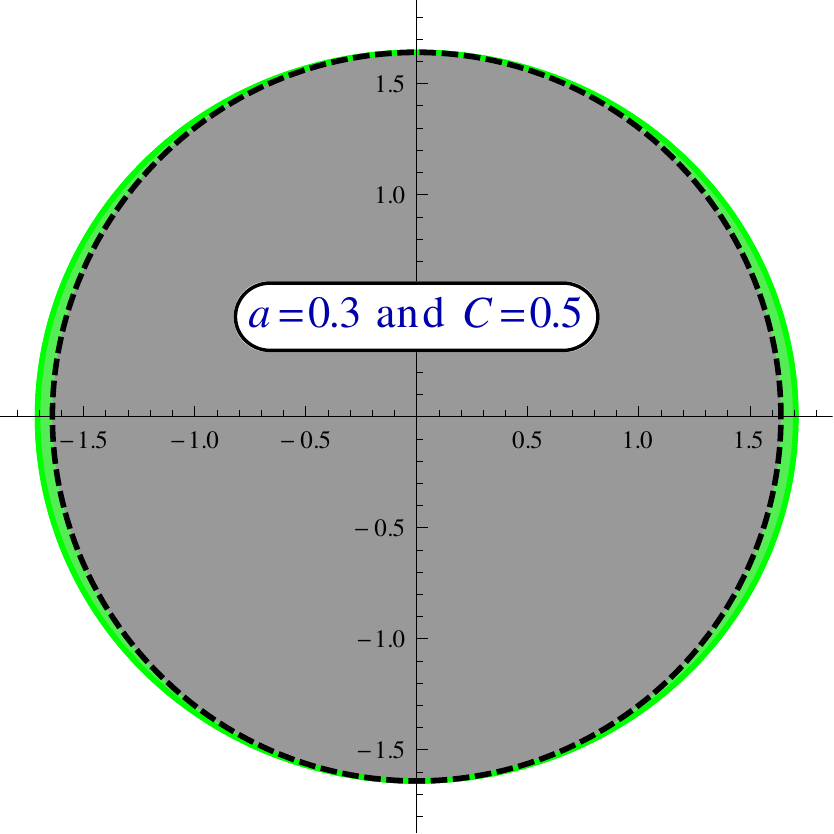}\hspace{0cm}
&\includegraphics[scale=0.66]{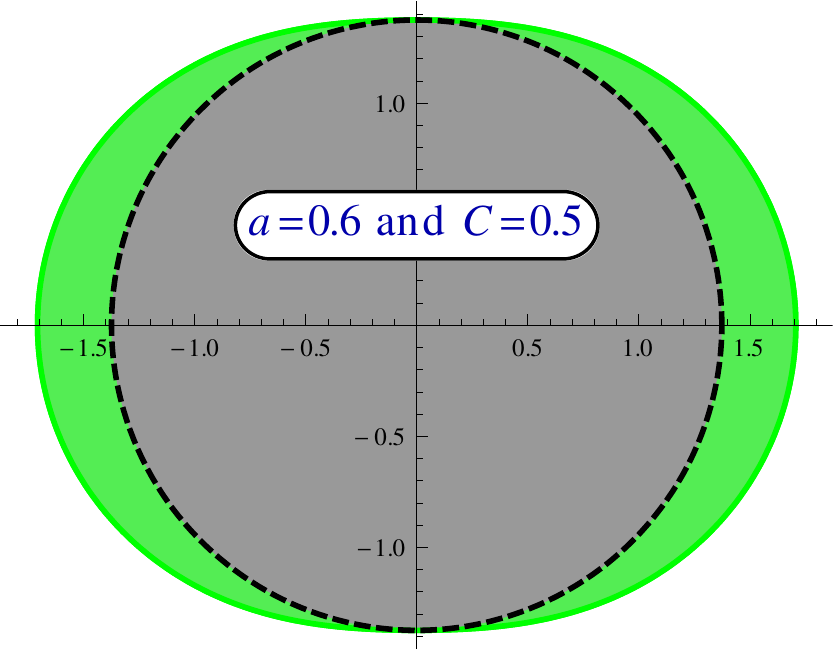}\hspace{0cm}
&\includegraphics[scale=0.66]{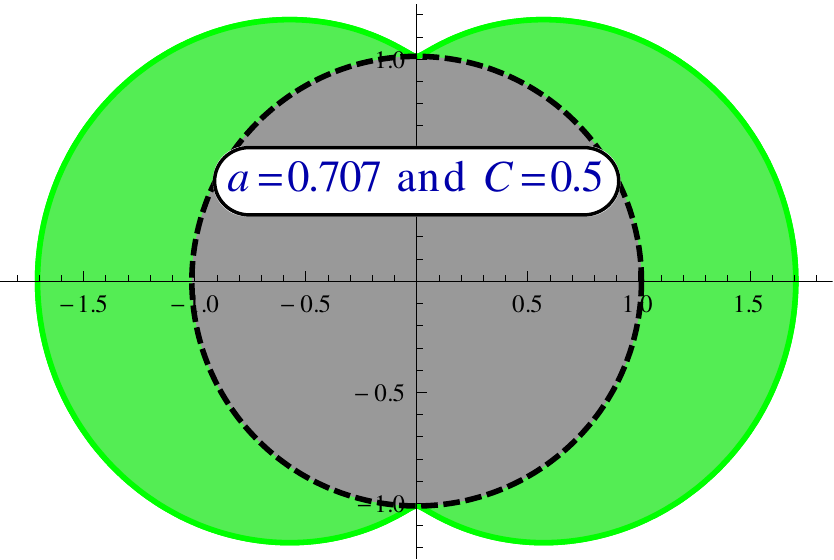}\hspace{0cm}
\\
\includegraphics[scale=0.66]{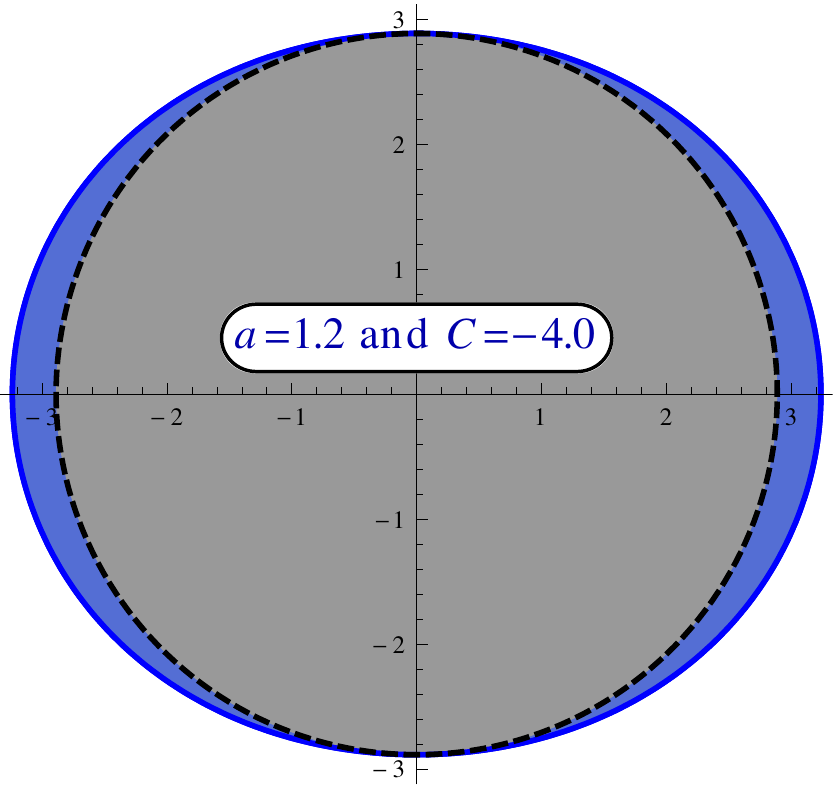}\hspace{0cm}
&\includegraphics[scale=0.66]{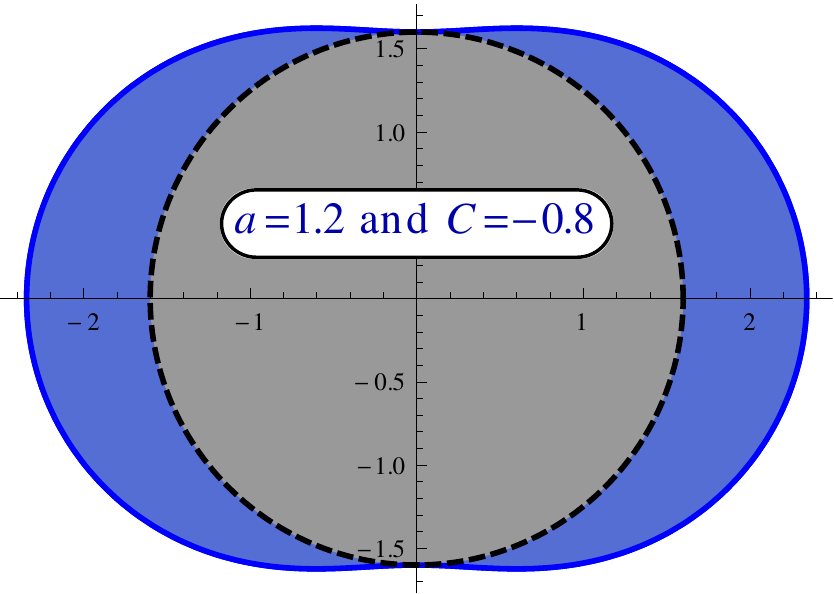}\hspace{0cm}
&\includegraphics[scale=0.66]{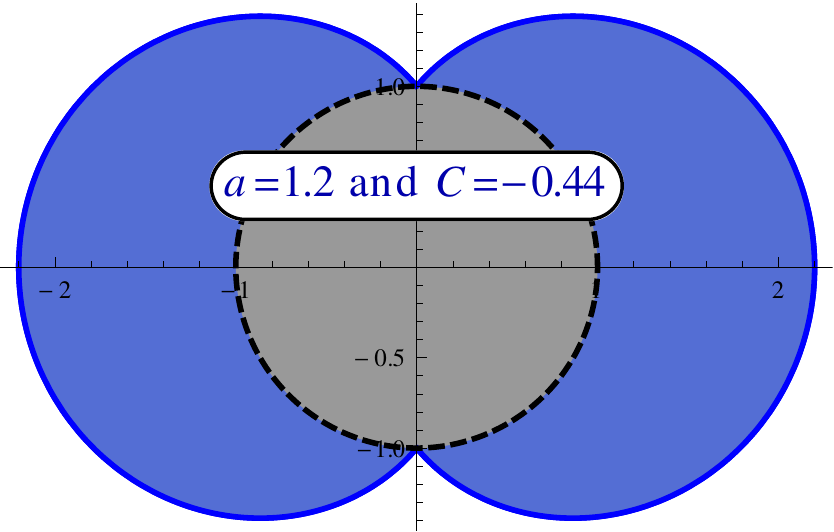}\hspace{0cm}
 \end{tabular}
 \caption{(Color-online) The shape of the ergosphere for different values of the rotation $a$ and the tidal charge parameters $\mathit{C}$ is plotted. Here, the solid (green and blue) lines represent the SLS, the dashed (black) lines represent the EH and we keep the value of the mass parameter $M=1$.}
 \label{fig1_ergo}
\end{figure*}
(EH) and the stationary limit surface (SLS). The EH is a null surface characterized by the radius $r_{EH}$ or $r_{+}$
determined by equating the contravariant component $g^{rr}$ to zero (i.e. $g^{rr}=0$),
\begin{equation}\label{horizon}
r_{\pm}= M\pm \sqrt{M^2-(a^2+\mathit{C})}.
\end{equation}

The event horizon does exist if $M^{2}\geq a^{2}+\mathit{C}$,
where the equality leads to the extremal BH case (i.e., $r_{+}=r_{-}$). Clearly, for the rotating braneworld BH the parameter $C$ can be negative \cite{Aliev:2005bi}.

The SLS (characterized by the radius $r_{SLS}$) is obtained by setting the prefactor of the term $dt^{2}$ in the metric equal to zero ($g_{tt}=0$).
The largest root of this equation gives the location of SLS around the BH
\begin{equation}\label{SLS}
r_{{SLS}}= M+\sqrt{M^2-(a^2 \cos^{2}{\theta}+\mathit{C})}.
\end{equation}

It is clear from Eq. (\ref{SLS}) that the SLS lies outside the EH except for the values $\theta=0$ and $\pi$ where it touches it. The region between the SLS and EH is known as the ergosphere. The negative value of tidal charge parameter $\mathit{C}$ for the case of braneworld-Kerr BH leads to the possibility of a greater EH and SLS in comparison with the
corresponding Kerr-Newman like BH where \textit{charge parameter} $Q^{2}=\mathit{C}>0$. It can be seen from the Eq. (\ref{horizon}) that when the parameter $\mathit{C}$ is negative, the EH do exists for extremal Kerr-Newman like braneworld BH if and only if the rotation parameter $a^{2}=M^{2}-\mathit{C}>M$. Hence, there exists a super-spinning case for the rotating braneworld BH and makes it more interesting in comparison with its Einstein-Maxwell theory counterpart (i.e., KN BH) \cite{Aliev:2005bi} where this situation is impossible. Using Eqs. (\ref{horizon}) and (\ref{SLS}), we found that the ergoregion lies within the limit $M<r<M+\sin\theta\sqrt{M^2-\mathit{C}}$ for the extremal case and from this, it is easy to conclude that the negative tidal charge parameter $\mathit{C}$ (i.e., $\mathit{C}<0$) of the rotating braneworld BH makes it more energetic than the usual Kerr-Newman BH which enables us to extract more rotational energy via the Penrose Process.

\begin{figure}[t]
\includegraphics[scale=0.75]{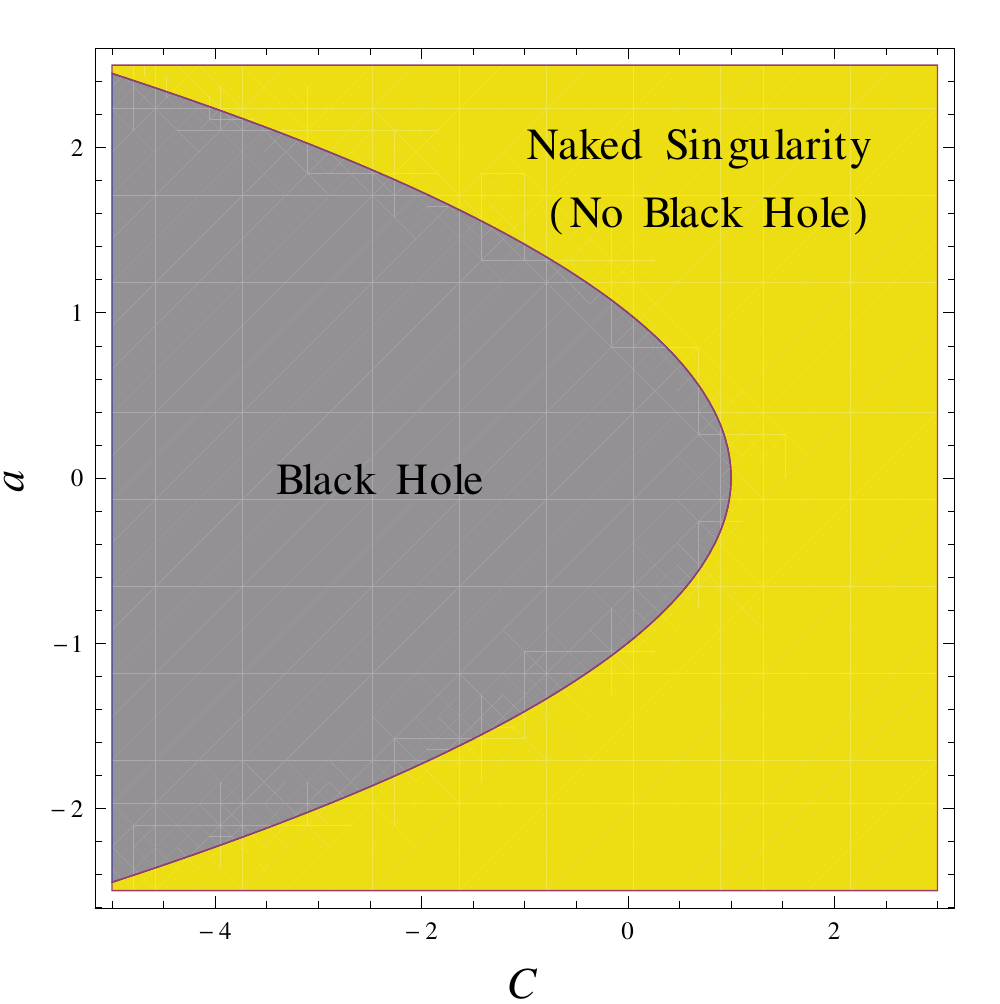}
\vskip 0mm \caption{(Color-online) The bounds on the rotation and the tidal charge parameters are shown for rotating braneworld BH.}
\label{fig2_cont}
\end{figure}

Fig. \ref{fig1_ergo} shows the effect of rotation and tidal charge parameters on the ergosphere. In each one of the first three rows of the figure, we fix the tidal charge parameter and vary the rotation parameter. The ergoregion (shown in green) in the first three rows increases with the increment in the rotation parameter. However, in the last row we fixed the rotation parameter $a>M$ and vary the tidal charge parameter. Here also, the ergoregion (shown in blue) increases as we increase the value of tidal charge for a fixed rotation parameter. Fig. \ref{fig2_cont} illustrates the bounds on the rotation and tidal charge parameters for the rotating braneworld BH. The gray region gives the values of parameters $a$ and $\mathit{C}$ for which we get an event horizon whereas the yellow region corresponds to the values of the same parameters for which there exists a naked singularity. The boundary of these two regions corresponds to the extremal rotating braneworld BH case.

\subsection{Conserved quantities}
Conserved quantities play a very important role in the study of a test body around a rotating BH as these help in simplifying the equations of motion. We have a conserved quantity $\mathcal{K}_{\psi}$, if there exists a Killing vector $\psi$ which satisfies the Killing equation $\psi_{\mu;\nu}+\psi_{\nu;\mu}=0$ (the semicolon means covariant derivative), the conserved quantity along the trajectory of a spinning particle is obtained with the help of master equation:
\begin{equation}\label{conserve_eq}
\mathcal{K}_{\psi}=P^{\mu}\psi_{\mu}-\frac{1}{2}S^{\mu \nu}\psi_{\mu;\nu}.
\end{equation}

Since we are dealing with stationary and axially symmetric spacetimes, there are two Killing vectors $\xi_{\mu}$ and $\Phi_{\mu}$, one time-like and the other space-like respectively (thus in addition to the mass $m$ and spin $S$ of a spinning particle, there exists two more conserved quantities).
These Killing vectors in a covariant tetrad base are given by
\begin{eqnarray}
\xi_{\mu}&=&-\left(\sqrt{\frac{\Delta}{\Sigma}}e^{(t)}_{\mu}+\frac{a\;\sin{\theta}}{\sqrt{\Sigma}}e^{(\phi)}_{\mu}\right),\label{timelike_killing}\\
\Phi_{\mu}&=&a\sqrt{\frac{\Delta}{\Sigma}}\;\sin^{2}{\theta}\;e^{(t)}_{\mu}+\frac{(r^{2}+a^{2})\;\sin{\theta}}{\sqrt{\Sigma}}e^{(\phi)}_{\mu}.
\label{spacelike_killing}
\end{eqnarray}
Here the tetrad frame of covariant vectors $e^{(b)}_{\mu}$ is given by
\begin{eqnarray}\label{tetrad2}
e^{(t)}_{\mu}&=&\left(\sqrt{\frac{\Delta}{\Sigma}},0,0,-a\;\sin^{2}{\theta}\sqrt{\frac{\Delta}{\Sigma}}\right),\nonumber\\
e^{(r)}_{\mu}&=&\left(0,\sqrt{\frac{\Delta}{\Sigma}},0,0,\right),\nonumber\\
e^{(\theta)}_{\mu}&=&\left(0,0,\sqrt{\Sigma},0,\right),\nonumber\\
e^{(\phi)}_{\mu}&=&\left(-\frac{a\;\sin{\theta}}{\sqrt{\Sigma}},0,0,\frac{(r^2+a^2)\;\text{sin}{\theta}}{\sqrt{\Sigma}}\right),
\end{eqnarray}
where the quantities within the bracket denote their tetrad components.

For convenience hereafter, we restrict ourselves to the equatorial plane ($\theta=\pi/2$) and use spin vector $S^{\mu}$ instead of spin tensor $S^{\mu \nu}$, satisfying the following relation in tetrad frame
\begin{eqnarray}\label{spin_vector}
S^{(b)}&=&-\frac{\epsilon^{(b)}_{(c)(d)(e)}V^{(c)}S^{(d)(e)}}{2m},\nonumber
\\
S^{(b)(c)}&=&m\epsilon^{(b)(c)}_{\;\;\;\;\;\;\;(d)(e)}V^{(d)}S^{(e)},
\end{eqnarray}
where, $\epsilon_{(b)(c)(d)(e)}$ is the completely antisymmetric tensor which satisfies the relation $\epsilon_{(t)(r)(\theta)(\phi)}=1$. As the spinning particle is restricted to the equatorial plane, the spin direction is always perpendicular to the plane $\theta=\pi/2$. Hence, $S^{(\theta)}$ is the only nonvanishing component of the spin vector $S^{(b)}$. For simplicity, we set $S^{(\theta)}\equiv-\bf{S}$. Here, the bold face $\bf{S}$ represents both the magnitude and the direction of a spin for the spinning particle. $\bf{S}>0$ represents that the direction of spin of a spinning particle is parallel to the rotation axis of the rotating braneworld BH, whereas $\bf{S}<0$ represents the antiparallel spin with respect to this axis. Furthermore, by using Eq. ({\ref{spin_vector}}), the nonzero tetrad components of the spin tensor $S^{(b)(c)}$ in terms of spin vector $S^{(b)}$  are
\begin{eqnarray}\label{spin_vector_equi}
S^{(t)(r)}&=&-m{\bf S}\;V^{(\phi)},\;\;
S^{(t)(\phi)}=m {\bf S}\;V^{(r)},\nonumber\\
&&S^{(r)(\phi)}=m{\bf S}\;V^{(t)}.
\end{eqnarray}

The conserved energy and the $z$ component of the total angular momentum $J_{z}$\footnote{It is worth to mention here that the conserved quantity representing $z$ component of the total angular momentum $J_{z}$ is sum of the orbital $L_{z}$ and the spin ${\bf S}$ angular  momenta.} of the spinning particle defined by using Eqs. (\ref{conserve_eq}), (\ref{timelike_killing}), (\ref{spacelike_killing}) and (\ref{spin_vector_equi}) read
\begin{eqnarray}\label{EL}
E&=&\frac{\mathcal{K_{\xi}}}{m}=\frac{1}{r}\left[\sqrt{\Delta}V^{(t)}+\left(a+\frac{{\bf S}}{r^{2}}(M r-\mathit{C})\right)V^{(\phi)}\right],\nonumber
\\
J_{z}&=&\frac{\mathcal{K_{\phi}}}{m}=\frac{1}{r}\Bigg[\sqrt{\Delta}(a+{\bf S})V^{(t)}+\bigg(r^2+a^2\nonumber\\
&& \;\;\;\;\;\;\;+\frac{a{\bf S}}{r}\left(r+M - \frac{\mathit{C}}{r}\right)\bigg)V^{(\phi)}\Bigg].
\end{eqnarray}

Using the Eqns. (\ref{uv}) and (\ref{EL}), we obtained the following nonvanishing components of four velocity in terms of four-momentum
\begin{equation}
U^{(t)}=\frac{A}{B}V^{(t)},\;\;U^{(r)}=\frac{A}{B}V^{(r)}\;\;\text{and}\;\;U^{(\phi)}=\frac{D}{B}V^{(\phi)},
\label{UV}
\end{equation}
where
\begin{eqnarray}
A&=&r^{4}+{\bf S}^{2}\left(\mathit{C}-M r\right),\nonumber\\
B&=&r^{4}+{\bf S}^{2}\left[(\mathit{C}-M r)+(4 \mathit{C}-3 M r)(V^{\phi})^{2}\right],\nonumber\\
D&=&r^{4}-{\bf S}^{2}(3\mathit{C}-2M r).
\label{ABC}
\end{eqnarray}

Next, with the help of tetrad relations (\ref{tetrad2}), the components of velocity fields for a spinning test particle in the background of a rotating braneworld BH can be written as
\begin{eqnarray}
\label{dtt}
\frac{dt}{d\tau}&=&\left(\frac{(a^{2}+r^{2})U^{(t)}+a \sqrt{\Delta}U^{(\phi)}}{r^{2}}\right)\sqrt{\frac{\Sigma}{\Delta}},\\
\label{dpt}
\frac{d\phi}{d\tau}&=&\left(\frac{a U^{(t)}+ \sqrt{\Delta}U^{(\phi)}}{r^{2}}\right)\sqrt{\frac{\Sigma}{\Delta}},\\
\label{dpr}
\frac{dr}{d\tau}&=&\frac{\sqrt{\Delta}}{r}U^{(r)}.
\end{eqnarray}

Now, by using the Eqns. (\ref{EL}), (\ref{UV}) and (\ref{ABC}), the explicit form of Eqns. (\ref{dtt})-(\ref{dpr}) reads
\begin{eqnarray}
&&\hspace{-0.5cm}\frac{dt}{d\tau}=\frac{a\left(1+\frac{{\bf S}^{2}\left(3 M-4\frac{\mathit{C}}{r}\right)}{r\sigma_{s}}\right)\left(J_{z}-E(a+{\bf S)}\right)+\frac{(r^{2}+a^{2})}{\Delta}P_{s}}{\sigma_{s}\Lambda_{s}},\nonumber\\
\label{dpt1}\\
&&\frac{d\phi}{d\tau}=\frac{\left(1+\frac{{\bf S}^{2}\left(3 M-4\frac{\mathit{C}}{r}\right)}{r\sigma_{s}}\right)\left(J_{z}-E(a+{\bf S)}\right)+\frac{a}{\Delta}P_{s}}{\sigma_{s}\Lambda_{s}},\nonumber\\
\label{dpp1}\\
&&\left(\frac{dr}{d\tau}\right)^{2}=\frac{R_{s}}{(\sigma_{s}\Lambda_{s})^{2}},\label{dpr1}
\end{eqnarray}
where
\begin{eqnarray}
\sigma_{s}&=&r^{2}\left[1-\frac{{\bf S^{2}}\left(M-\frac{\mathit{C}}{r}\right)}{r^{3}}\right],\nonumber\\
\Lambda_{s}&=&\left[1+\frac{{\bf S}^2(4\mathit{C}-3M r)(J_{z}-E(a+s))^{2}}{\sigma_{s}^{3}}\right],\nonumber\\
P_{s}&=&\left[r^{2}+a^{2}+\frac{a{\bf S}}{r}\left(r+M-\frac{
\mathit{C}}{r}\right)\right]E\nonumber\\
&&-\left[a+\frac{{\bf S}(M r-\mathit{C})}{r^2}\right]J_{z}\nonumber\\
R_{s}&=&\left[P_{s}^2-\Delta\left(\left(\frac{\sigma_{s}}{r}\right)^2+(J_{z}-(a+{\bf S})E)^2\right)\right].
\end{eqnarray}

\section{Effective potential}
\label{Effective_potential}
Analyzing the effective potential is important for studying the dynamics of the spinning particle moving in the vicinity of a BH. Hence, in this section we wish to bring out the effect of rotation and tidal charge parameters on the effective potential of the rotating braneworld BH. We consider equatorial motion solely, taking into account only the equation for the radial velocity. We rewrite the Eq. (\ref{dpr1}) in quadratic form for the parameter $E$ as
\begin{equation}
\Upsilon E^{2}-2\;\Xi\; E+\Pi-\left[\sigma_{s}\Lambda_{s}\left(\frac{dr}{d\tau}\right)\right]^{2}=0,
\label{E2}
\end{equation}
\begin{figure*}
\begin{tabular}{c c c c c}
\hspace{-0.7cm}
\includegraphics[scale=0.51]{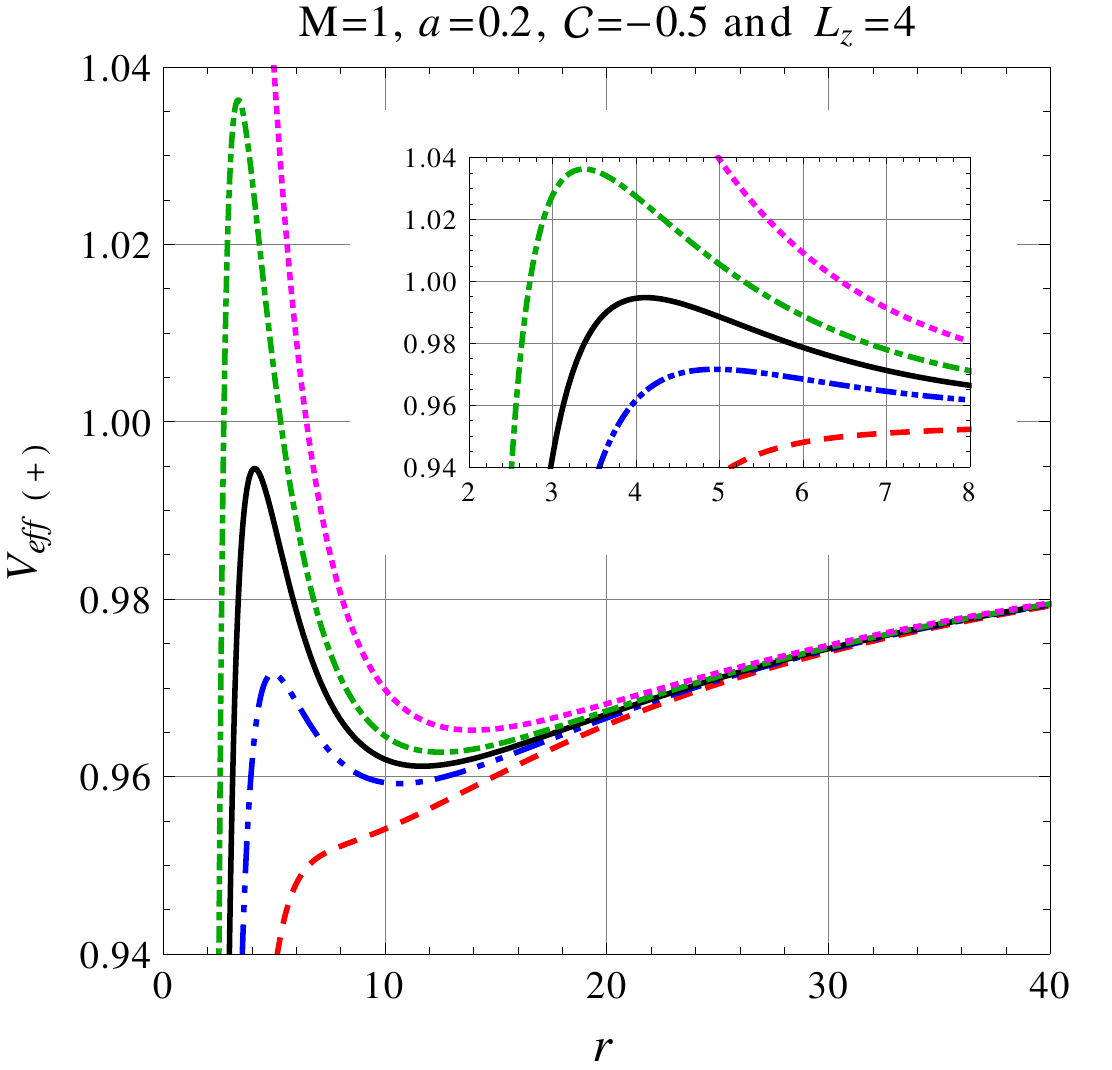}\hspace{-0.3cm}
&\includegraphics[scale=0.51]{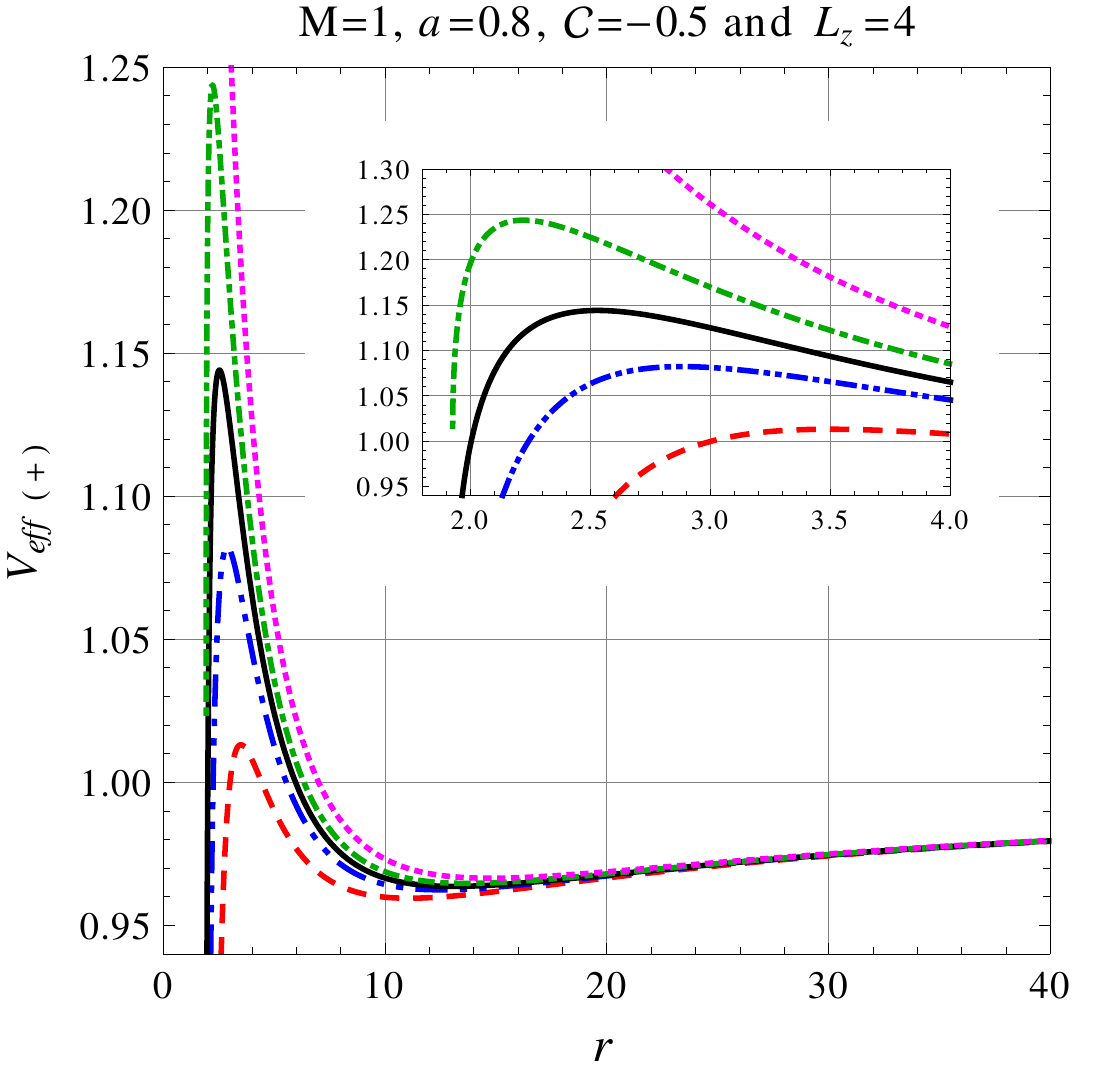}\hspace{-0.3cm}
&\includegraphics[scale=0.51]{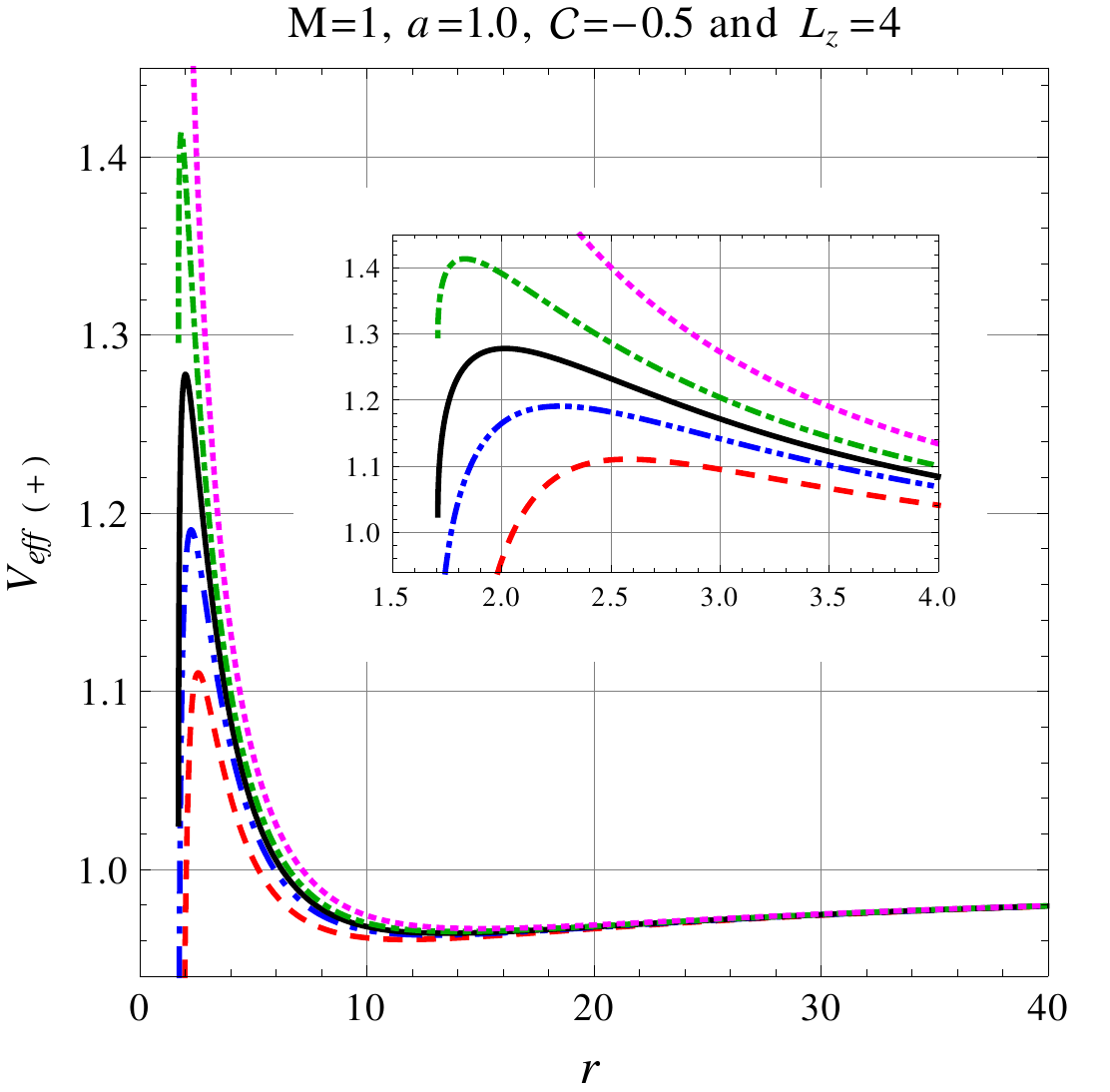}
\\
\hspace{-0.7cm}
\includegraphics[scale=0.51]{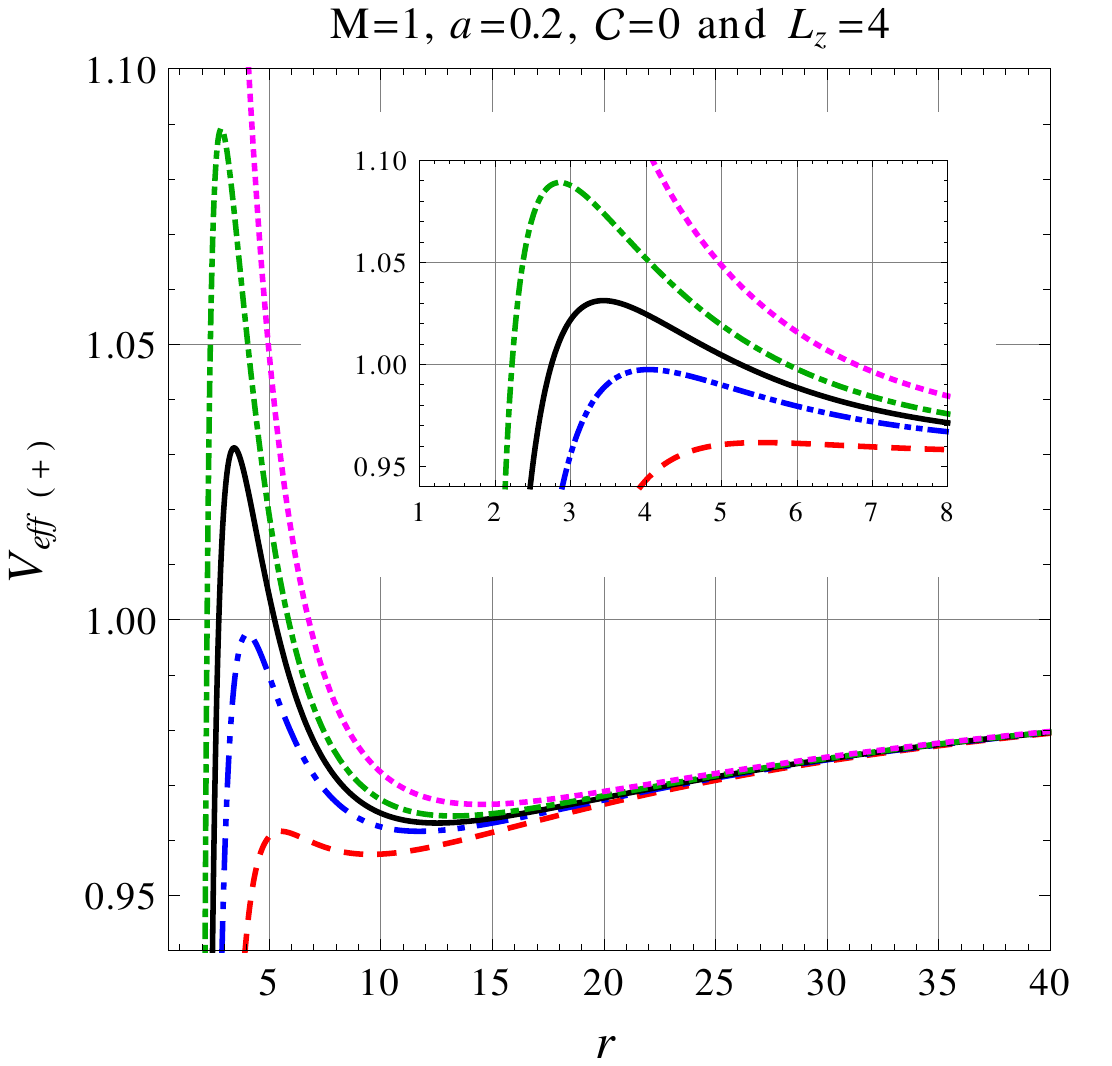}\hspace{-0.3cm}
&\includegraphics[scale=0.51]{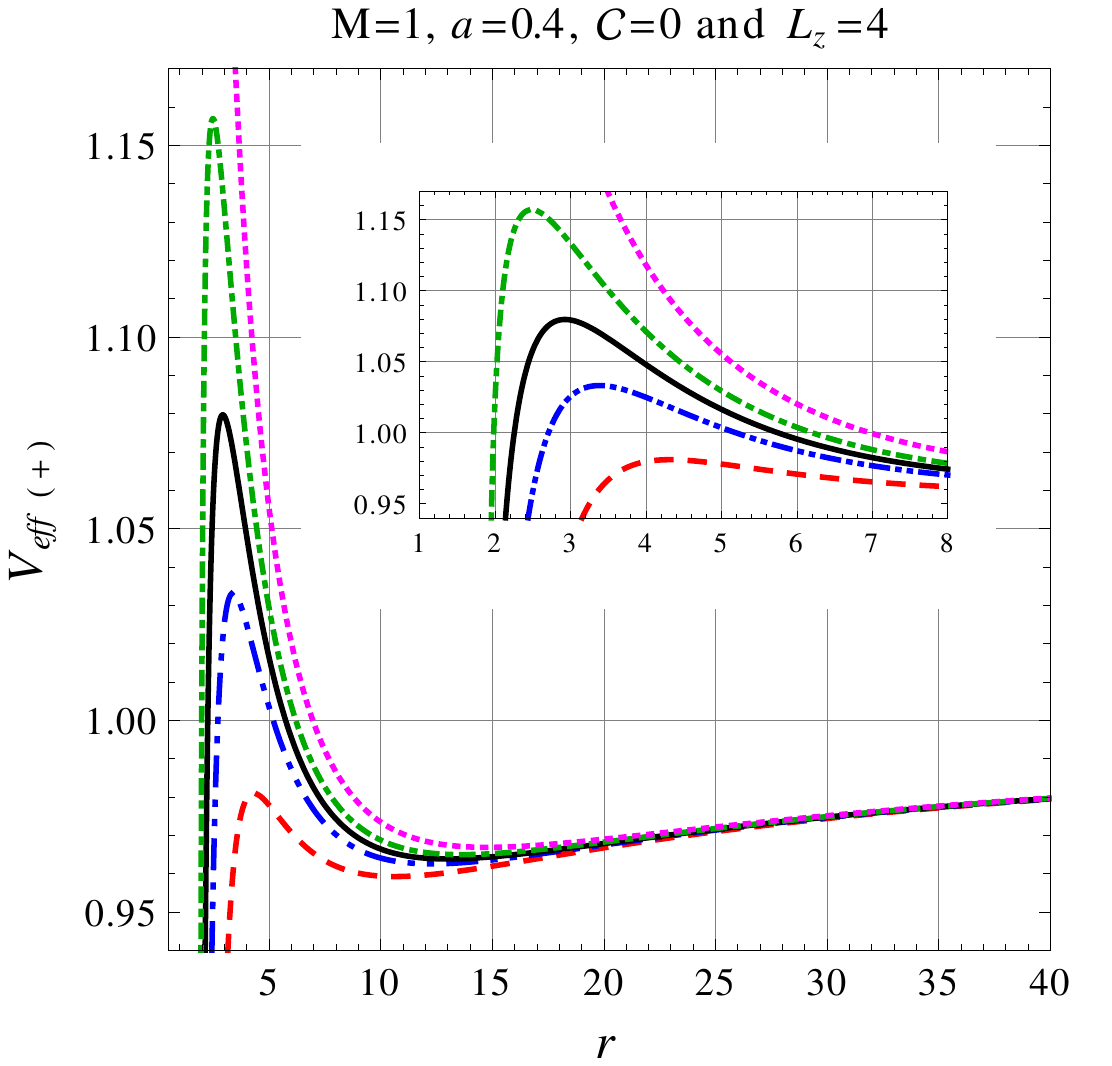}\hspace{-0.3cm}
&\includegraphics[scale=0.51]{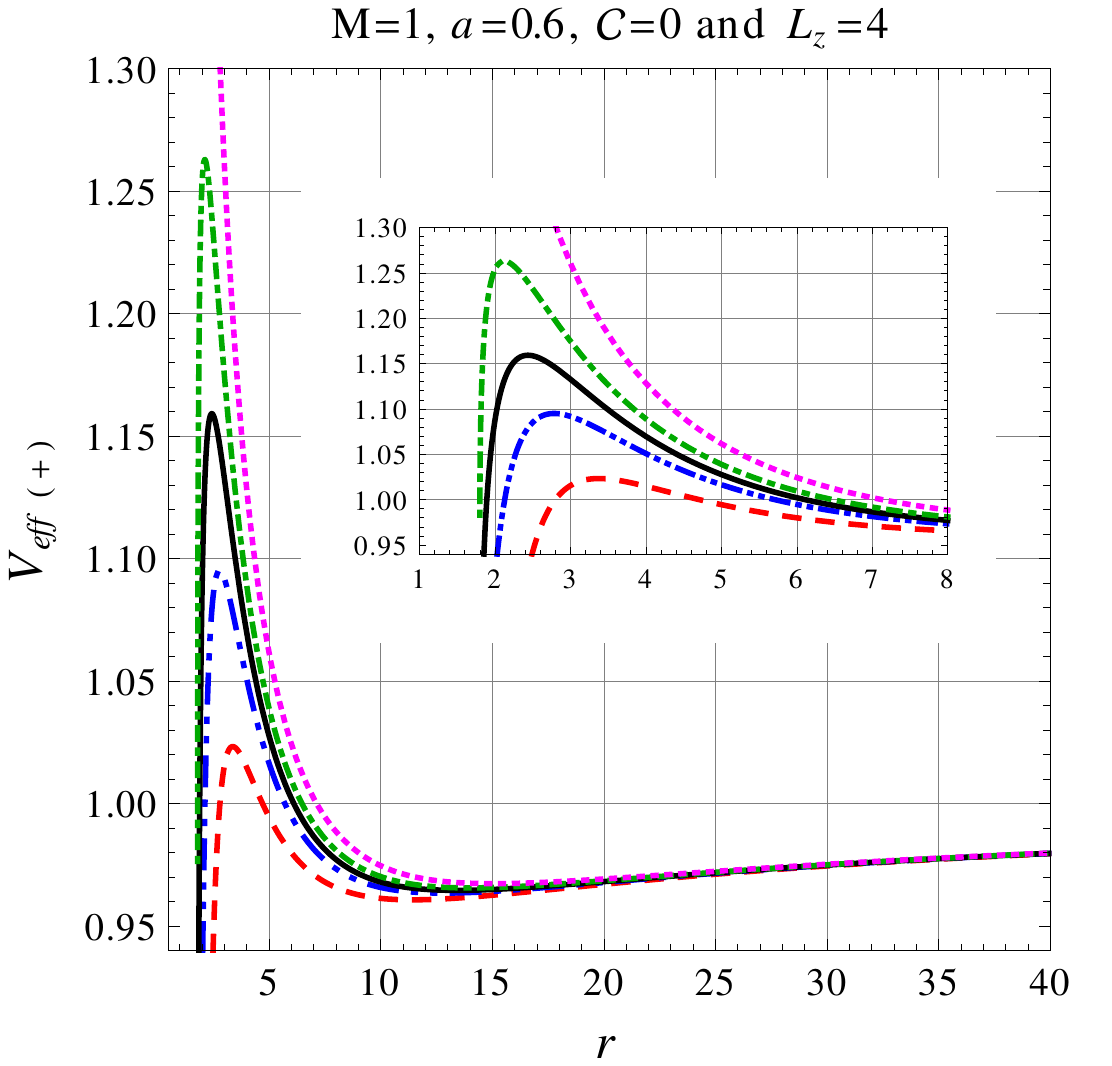}
\\
\hspace{-0.7cm}
\includegraphics[scale=0.51]{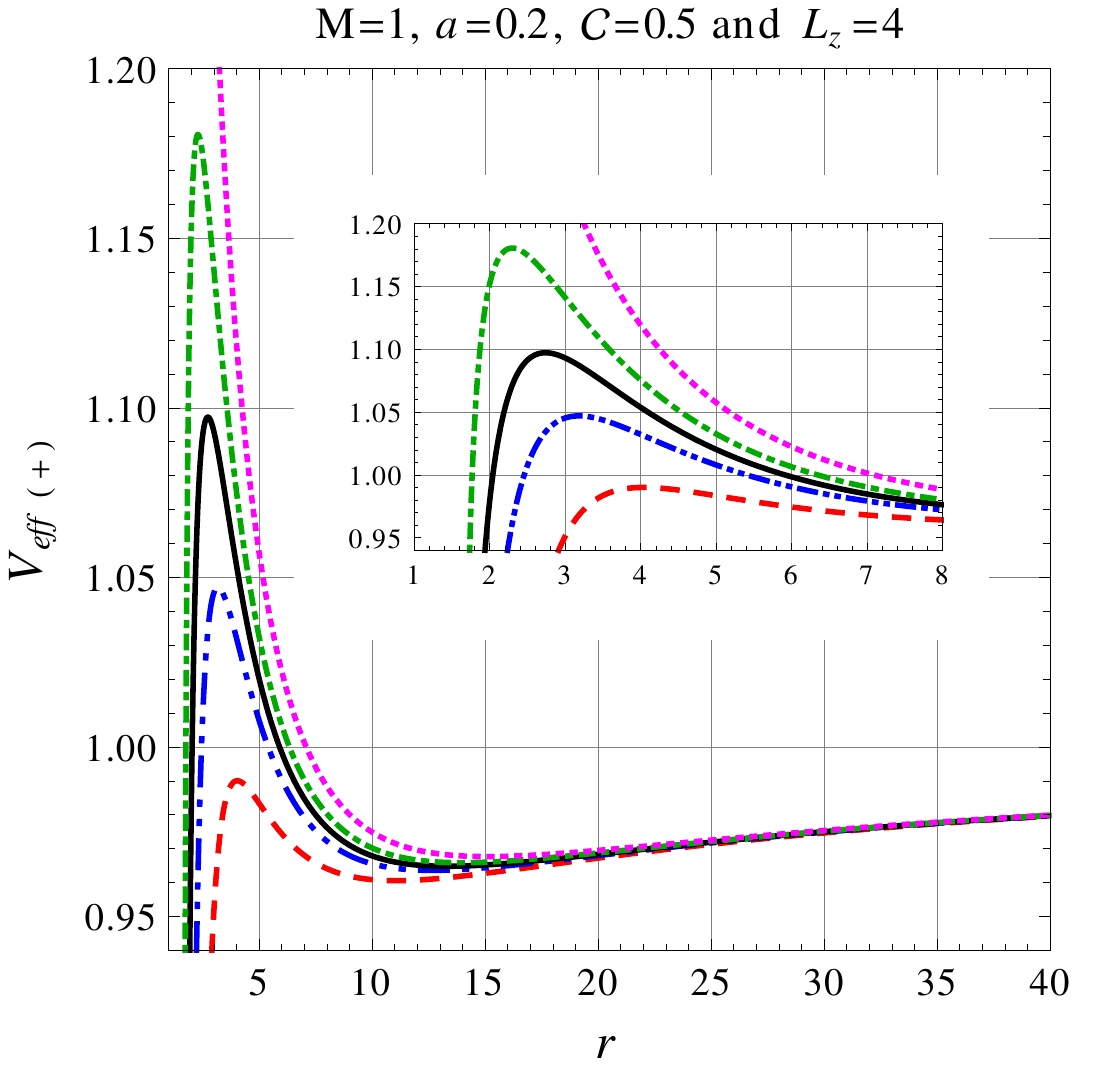}\hspace{-0.3cm}
&\includegraphics[scale=0.51]{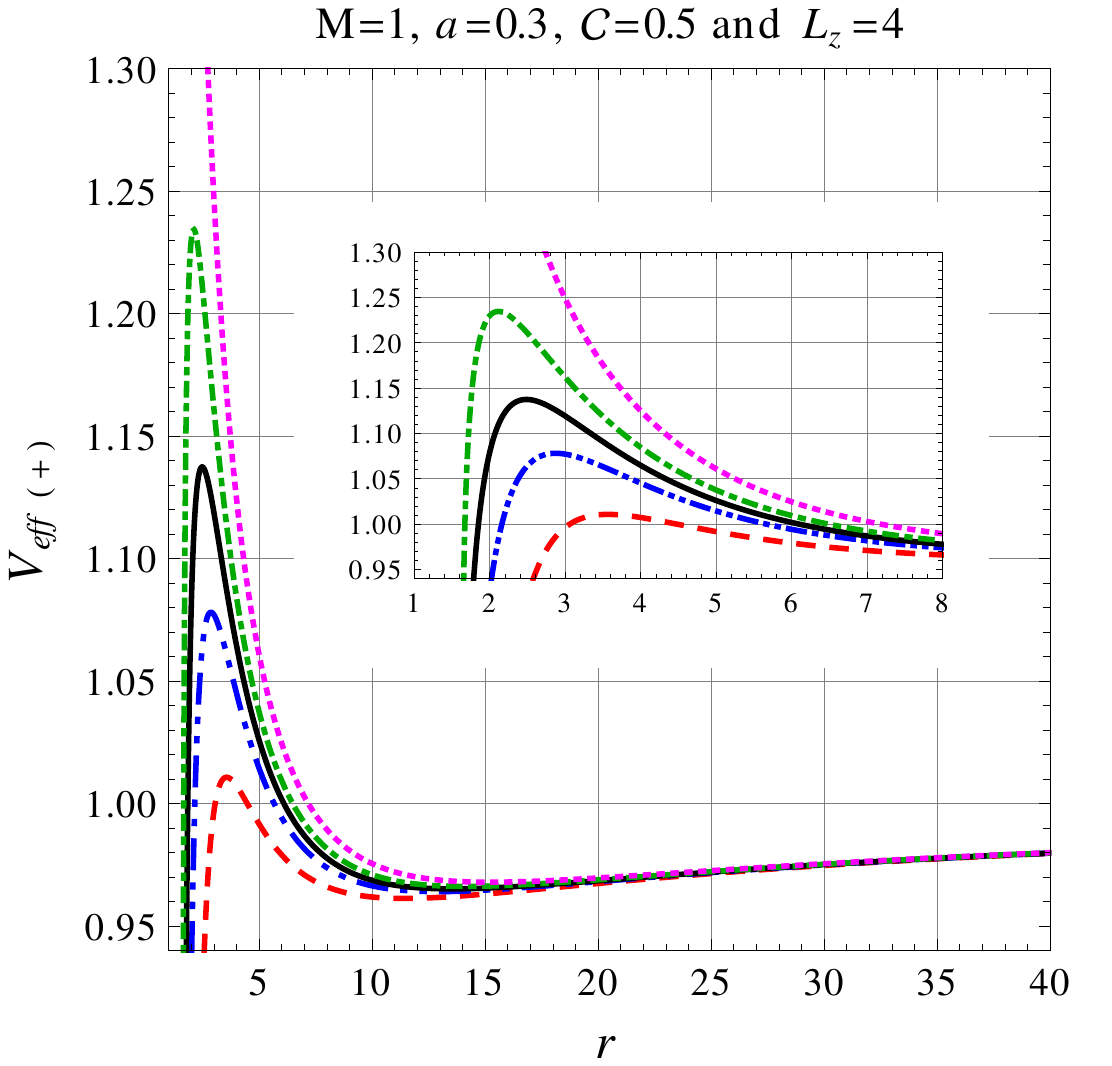}\hspace{-0.3cm}
&\includegraphics[scale=0.51]{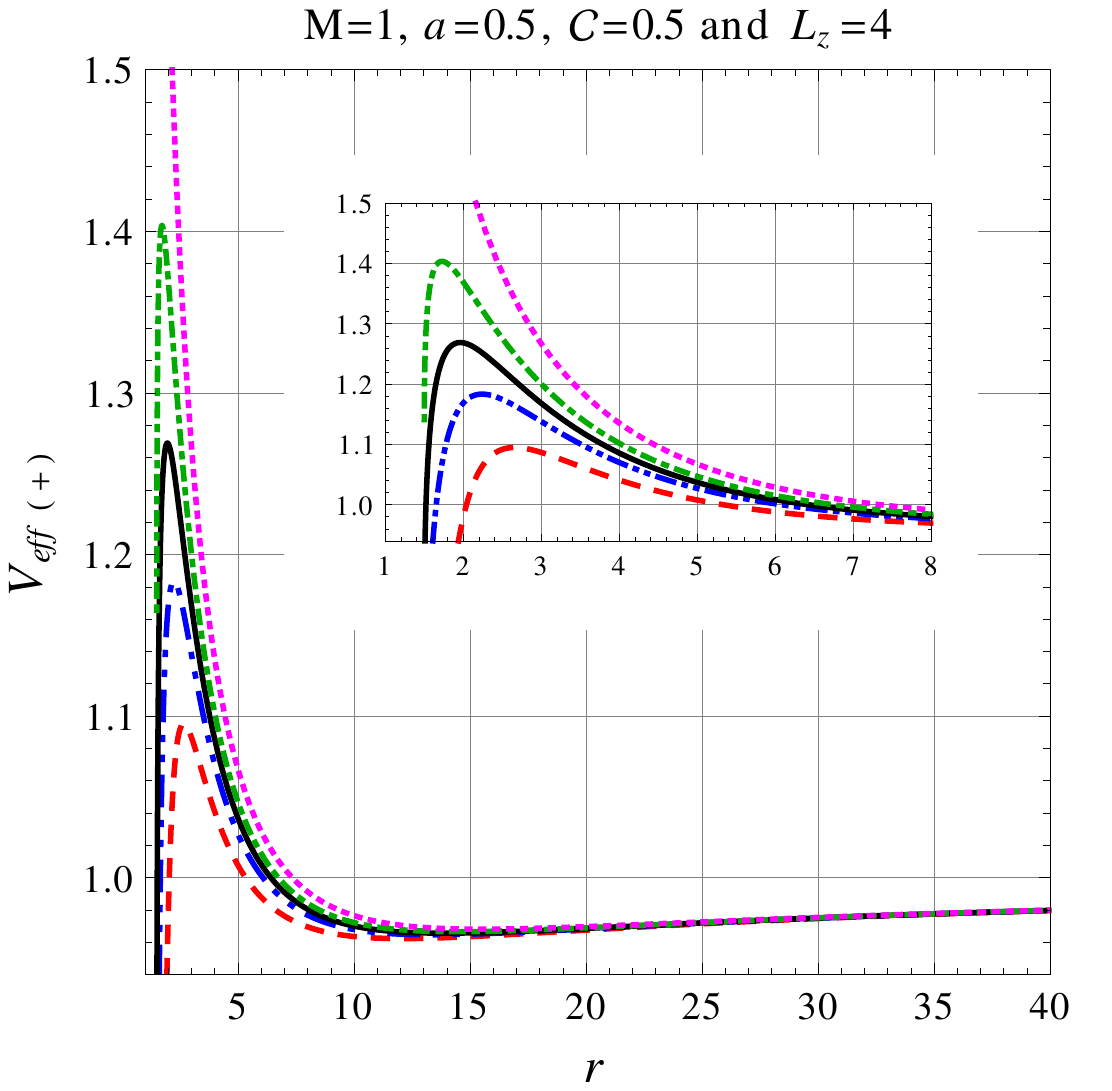}
\\
\hspace{-0.7cm}
\includegraphics[scale=0.51]{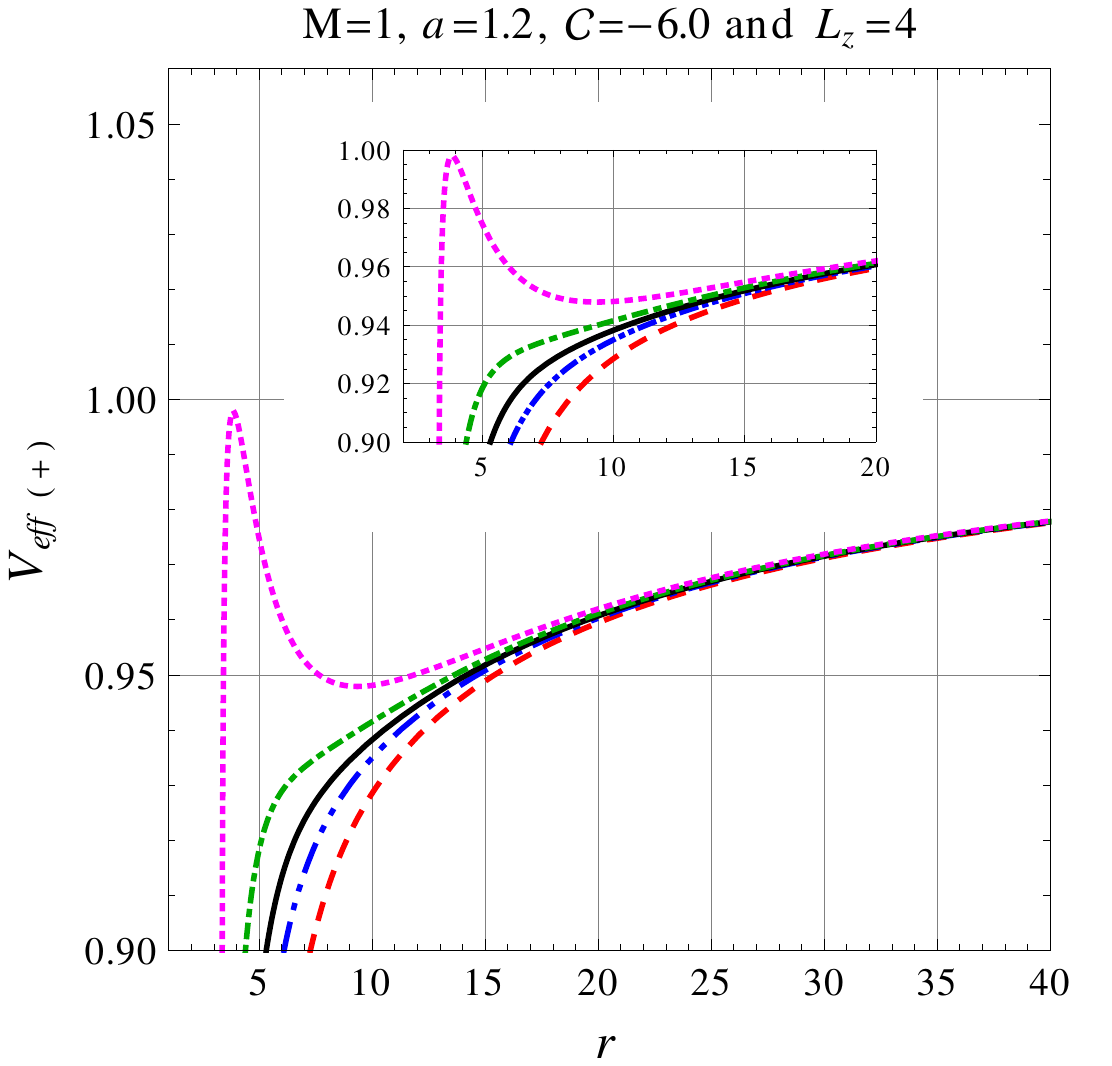}\hspace{-0.3cm}
&\includegraphics[scale=0.51]{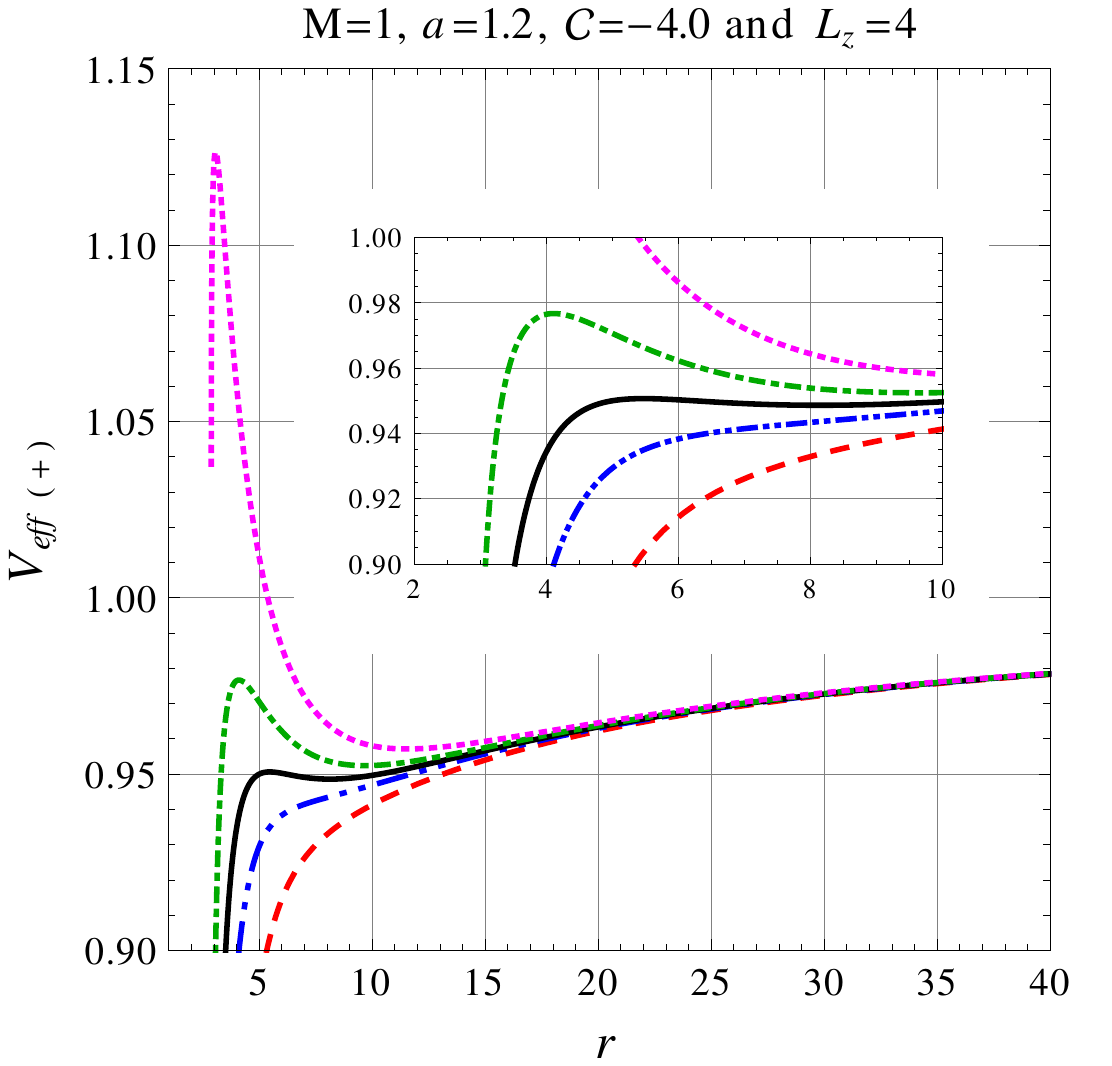}\hspace{-0.3cm}
&\includegraphics[scale=0.51]{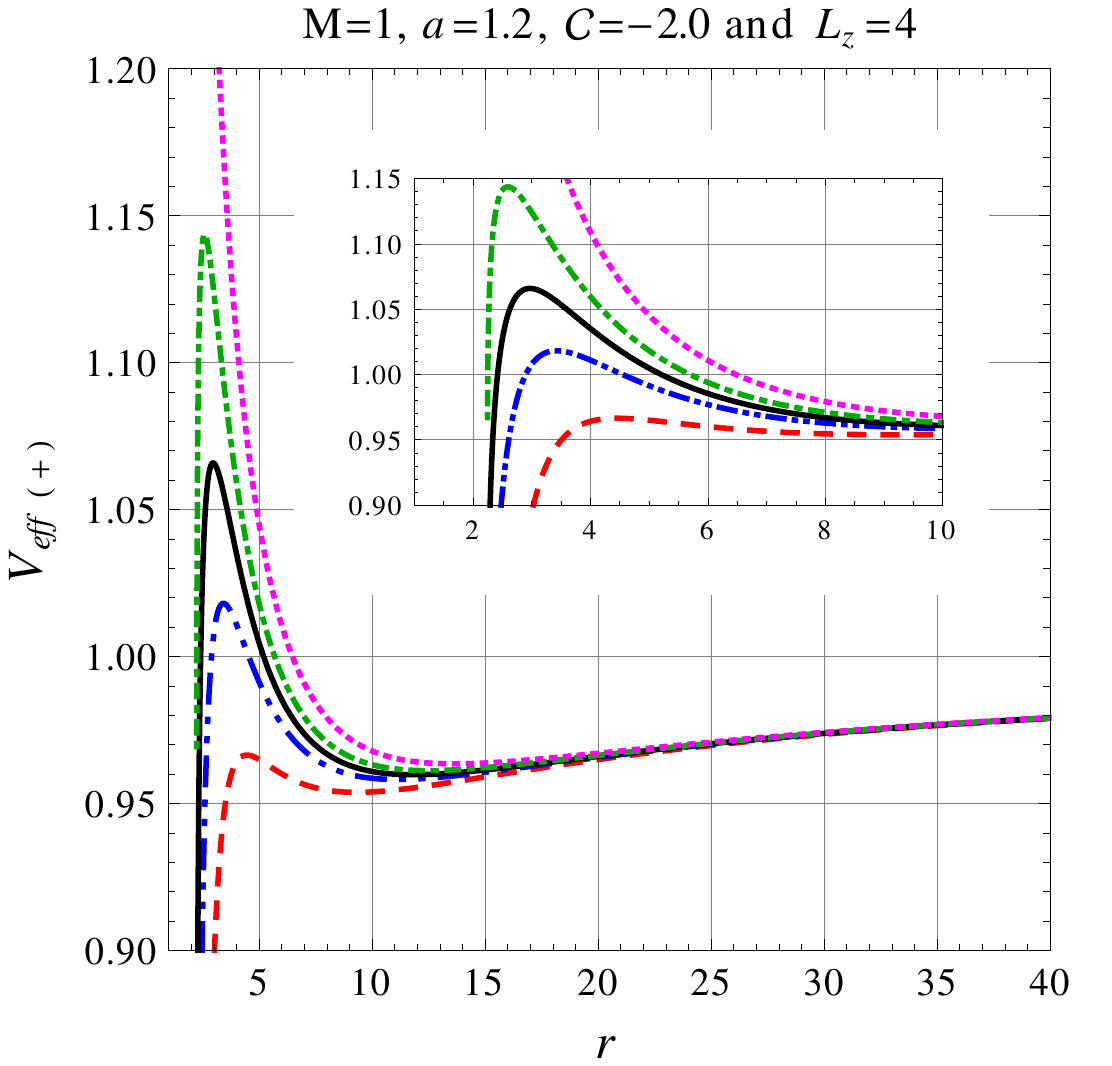}
\end{tabular}
 \caption{(Color-online) Plots of the effective potential of a spinning test particle moving in the rotating braneworld BH background for different values of particle spin (${\bf S}$). Here, for dashed (Red) curve ${\bf S}=-1.5$, for double dot dashed (Blue) curve ${\bf S}=-0.5$, for solid (Black) curve ${\bf S}=0$, for dot dashed (Green) curve ${\bf S}=0.5$, for dotted (Magenta) curve ${\bf S}=1.5$ and the mass parameter $M=1$.}
 \label{fig3_veff}
\end{figure*}
where,
\begin{eqnarray}
\Upsilon&\equiv&\left[r^{2}+a^{2}+\frac{a {\bf S}}{r}\left(r+M-\frac{\mathit{C}}{r}\right)\right]^{2}-\Delta (a+{\bf S})^{2},\nonumber\\
\Xi&\equiv&\left[\left(a+\frac{{\bf S}\left(M-\frac{\mathit{C}}{r}\right)}{r}\right)\left(r^{2}+a^{2}+\frac{(a{\bf S})\left(r+M-\frac{\mathit{C}}{r}\right)}{r}\right)\right.\nonumber\\
&& - \Delta(a+{\bf S})\Bigg]J_{z},\nonumber\\
\Pi&\equiv& \left[a+\frac{{\bf S}\left(M-\frac{\mathit{C}}{r}\right)}{r}\right]J_{z}^{2}-\Delta\left[\left(\frac{\sigma}{r}\right)^{2}+J_{z}^{2}\right].
\end{eqnarray}

Eqn. (\ref{E2}) can also be written in the form
\begin{equation}
(E-V_{eff(+)})(E-V_{eff(-)})-\left[\sigma_{s}\Lambda_{s}\left(\frac{dr}{d\tau}\right)\right]^{2}=0,
\end{equation}
where
\begin{equation}
V_{eff(\pm)}\equiv\frac{\Xi\pm \left(\Xi^{2}+\Upsilon \Pi\right)^{\frac{1}{2}}}{\Upsilon}.
\end{equation}

Hereafter, we work only with $V_{eff(+)}$ because this is the physical effective potential corresponding to the spinning particles with future pointing four momentum \cite{MTW}. For the above stated reason in Fig. \ref{fig3_veff}, we only plot $V_{eff(+)}$ for some values of parameters ($a, L_{z}$, $\mathit{C}$ and ${\bf S}$). The effective potential $V_{eff(+)}$ of the spinning particle can have one or two extreme points as shown in Fig. \ref{fig3_veff}, where it has circular orbits. The extreme point corresponding to maximum value of $V_{eff(+)}$ leads to unstable circular orbits while the extreme point corresponding to its minimum value leads to stable circular orbits. From Fig. \ref{fig3_veff}, it is evident that $V_{eff(+)}$ is very sensitive to the test particle's spin ${\bf S}$ because the maximum and the minimum values of $V_{eff(+)}$ increases with the rise of ${\bf S}$. In Fig. \ref{fig3_veff}, we fixed the tidal charge $\mathit{C}$ parameter and varied the rotation parameter $a$ along each row except the last row where parameter $a$ is fixed and parameter $\mathit{C}$ varies. It is clear from each of the first three rows of Fig. \ref{fig3_veff} that the unstable circular orbits shift closer to the EH of rotating braneworld BH as the rotation parameter $a$ goes higher for the respective value of ${\bf S}$. Additionally, we notice a similar kind of behavior in the last row, when the $a$ is fixed and the parameter $\mathit{C}$ grows from left to right along the row.

\section{Inner most stable circular orbit (ISCO) and the superluminal constraint}
\label{ISCO}
In this section, we numerically study the behavior of ISCO parameters (i.e., $r, E$ and $L_{z}$) for different values of rotation $a$ and tidal charge $\mathit{C}$ parameters as a function of the test particle spin {\bf S}. It is known from literature \cite{Suzuki:1997by,Jefremov:2015gza,Lukes-Gerakopoulos:2017vkj,Zhang:2017nhl} that to investigate the ISCO of the spinning test particle in a curved background we need to solve a system which is comprising of the following three conditions:\\
\begin{eqnarray}
&&\frac{dr}{d\tau}=0\;\Rightarrow\; \left(\frac{dr}{d\tau}\right)^{2}\equiv V_{S}=0\;,\label{radial_v}\\
&&\frac{d^{2}r}{d\tau^{2}}=0\;\Rightarrow\;\frac{dV_{S}}{dr}=0\;,\label{radial_a}\\
&&\frac{d^{2}V_{S}}{dr^{2}}=0\; \label{ISCO_cond}.
\end{eqnarray}

Here, the Eq. (\ref{radial_v}) signifies that the radial velocity of spinning particle is zero and Eq. (\ref{radial_a}) means that the radial velocity for the spinning particle is constant, which means, in other words, that the radial acceleration should vanish. If both Eqs. (\ref{radial_v}) and (\ref{radial_a}) hold true simultaneously, the spinning particle will move in a circular orbit around the rotating braneworld BH. Now, in order to find the location of the ISCO, the point where the maximum and minimum of the $V_{eff(+)}$ meets, we need to introduce one more condition (i.e., Eq. (\ref{ISCO_cond})) in addition to the conditions described by Eqs. (\ref{radial_v}) and (\ref{radial_a}). The explicit form of these three equations are shown in Appendix is used to find the behavior of the ISCO 
\begin{figure}[h!]
\begin{tabular}{c}
{(a)}\includegraphics[scale=0.89]{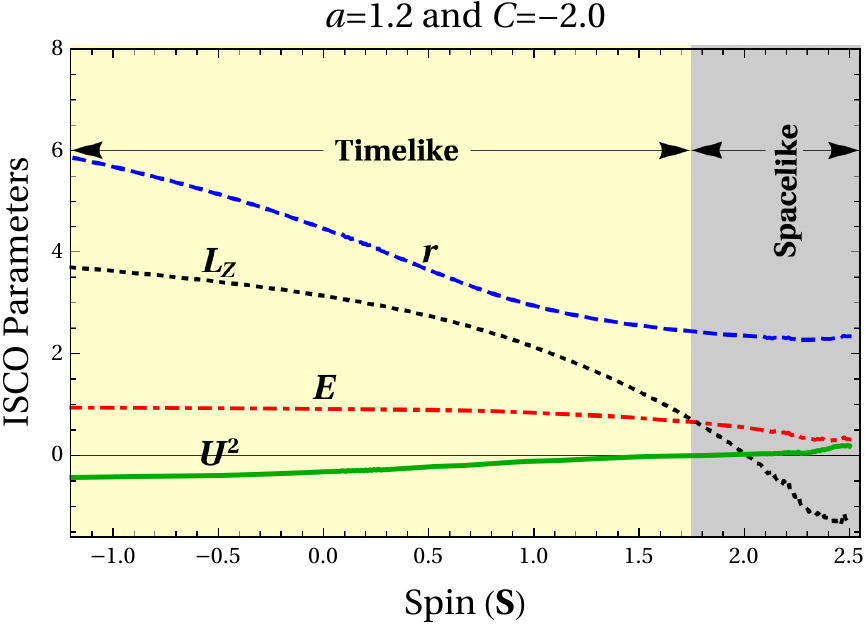}\hspace{0cm}
\\
{(b)}\includegraphics[scale=0.89]{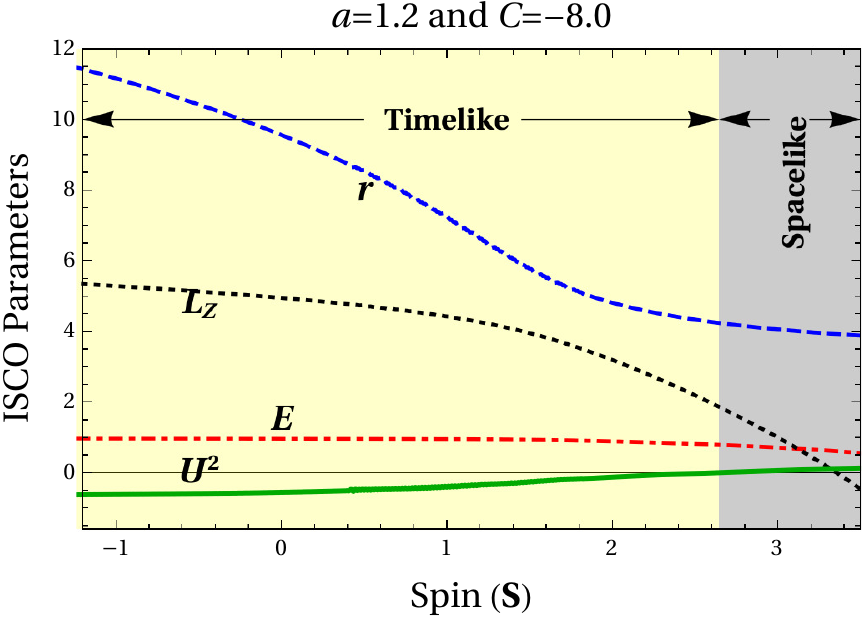}\hspace{0cm}
\end{tabular}
 \caption{(Color-online) Behavior of ISCO parameters ($r, E$ and $L_{z}$) and superluminal constraint ($U^{2}$) as a function of spin ($S$) for the spinning test particle in  the rotating braneworld BH for corotating case ($J_{z}>0$). Here also, we keep the mass parameter $M$ equals to unity.}
\label{fig4_ISCO_parameters_co_rotating}
\end{figure}
parameters as a function of the particle spin ${\bf S}$. It is worth mentioning here that for the case of spinning particle, the four-velocity and the four- momentum are not parallel and hence the four velocity can be timelike ($U^{2}<0$) or spacelike ($U^{2}>0$), where $U^{2}\equiv U_{\mu}U^{\mu}$. Therefore, to study the behavior of the ISCO parameters as the function of the parameter ${\bf S}$ we need to take into consideration the superluminal constraint as well,
\begin{equation}
\label{super_luminal_condition}
\frac{U^{2}}{(U^{t})^{2}}=
g_{tt}+g_{rr}\left(\frac{dr}{dt}\right)^{2}+
g_{\phi\phi}\left(\frac{d\phi}{dt}\right)^{2}+
2g_{t\phi}\left(\frac{d\phi}{dt}\right)<0 ,
\end{equation}
which gives information of the region where the circular motion of the spinning particle will be superluminal (unphysical behavior) or subluminal (physical behavior).

\begin{figure}[ht!]
\begin{tabular}{c}
{(a)}\includegraphics[scale=0.890]{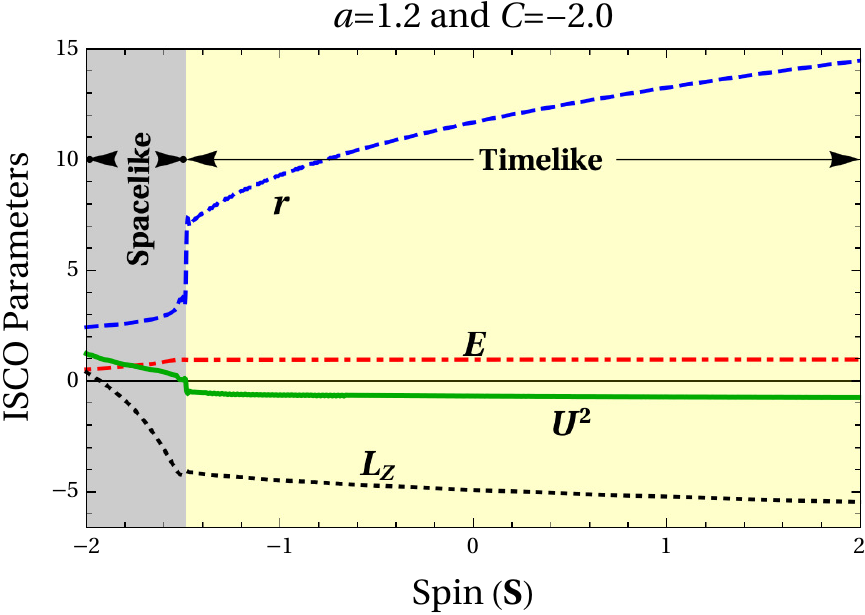}\hspace{0cm}
\\
{(b)}\includegraphics[scale=0.890]{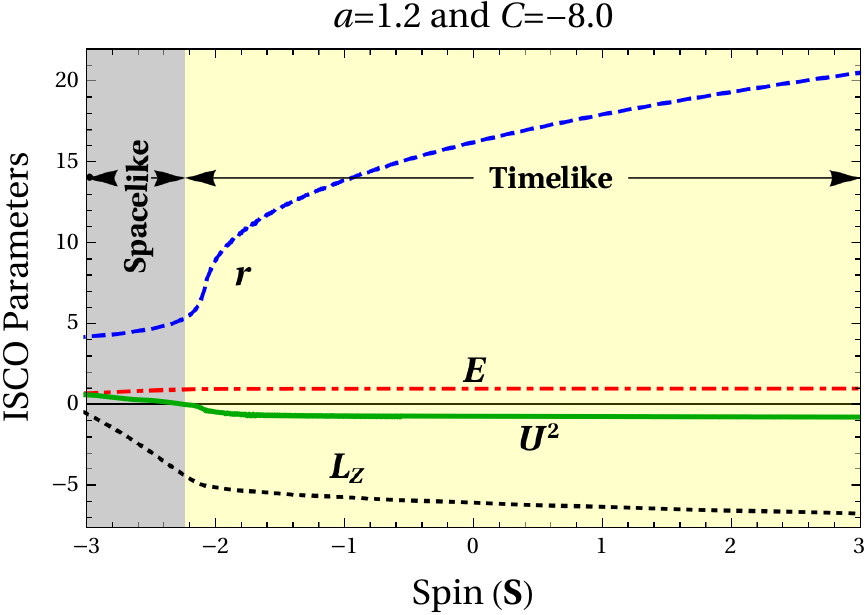}\hspace{0cm}
\end{tabular}
 \caption{(Color-online) Behavior of ISCO parameters ($r, E$ and $L_{z}$) and superluminal constraint ($U^{2}$) as a function of spin ($S$) for the spinning test particle in the rotating braneworld BH for counter-rotating ($J_{z}<0$) cases. Here, $M$ equals to unity.}
\label{fig5_ISCO_parameters_counter_rotating}
\end{figure}

In Figs. \ref{fig4_ISCO_parameters_co_rotating} and \ref{fig5_ISCO_parameters_counter_rotating}, we numerically explore the behavior of ISCO parameters ($r, L_{z}$ and $E$) together with the superluminal constraint as a function of the particle spin ${\bf S}$ for both corotating ($J_{z}>0$) and counter-rotating ($J_{z}<0$) cases in the rotating braneworld BH. We divide Figs. \ref{fig4_ISCO_parameters_co_rotating} and \ref{fig5_ISCO_parameters_counter_rotating} into two parts to bring out the effect of tidal charge parameter $\mathit{C}$ on the ISCO parameters. In both Figs. \ref{fig4_ISCO_parameters_co_rotating} and \ref{fig5_ISCO_parameters_counter_rotating}, the parts (a) and (b) correspond to the parameter $\mathit{C}=-2.0$ and $-8.0$, respectively for fixed value of the parameter $a=1.2$. With the help of Figs. \ref{fig4_ISCO_parameters_co_rotating}, \ref{fig5_ISCO_parameters_counter_rotating}, \ref{fig6_Phase_Plot1} and \ref{fig7_Phase_Plot2}, a brief summary of the results about the behavior of ISCO parameters($r, L_{z}$ and $E$), superluminal constraint (\ref{super_luminal_condition}) and the particle spin ${\bf S}$ is presented as follows:
\begin{itemize}
\item For the Kerr-Newman like braneworld BH case (i.e. $a>0$ and $\mathit{C}<0$), the physical values of ISCO parameters decrease as the spin ${\bf S}$ of the particle increases for the corotating cases, whereas for the counter-rotating cases, the ISCO parameter $r$ increases and the orbital angular momentum parameter $L_{z}$ decreases with the increases in the spin ${\bf S}$ of the particle.
\begin{figure}
\begin{tabular}{c}
\includegraphics[scale=0.89]{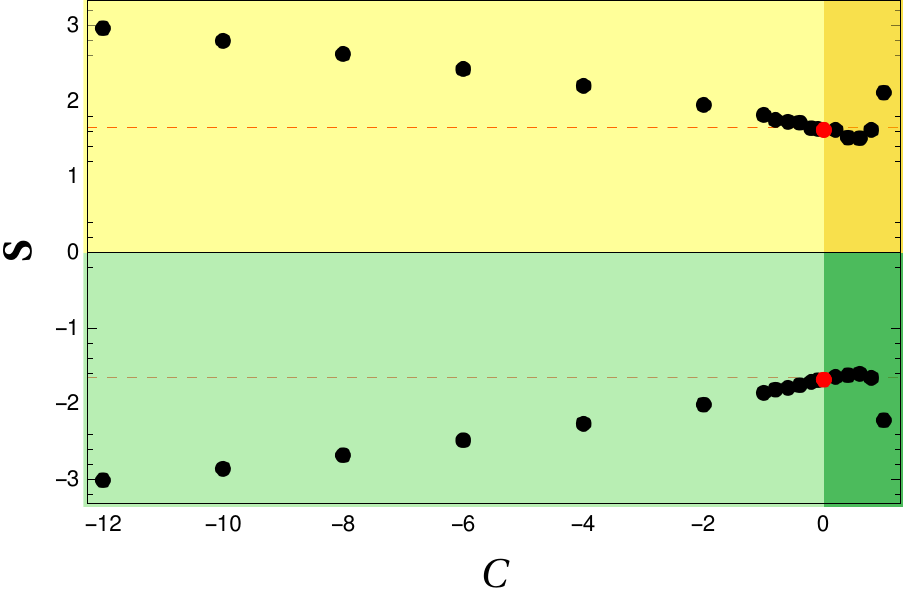}\hspace{0cm}
\end{tabular}
 \caption{(Color-online) Phase plot between the parameters ${\bf S}$ and $C$, shows the bound on the particle's spin ${\bf S}$ for different set of values of parameter $C$ in the case $a=0$.
 The permitted region is between the curves formed by the dots.
 Here, light and dark yellow regions represent corotating ($J_{z}>0$) cases of the spinning particle around the Reissner-Nordstr$\ddot{o}$m like braneworld BH ($\mathit{C}<0$) and the Reissner-Nordstr$\ddot{o}$m like BH ($\mathit{C}>0$), respectively. The same applies for the green region representing the counter-rotating ($J_{z}<0$) cases. The red-dot in the yellow and green regions shows the limiting value of parameter ${\bf S}$ for corotating and counter-rotating ($J_{z}<0$) ISCOs around Schwarzschild BH. The mass parameter $M=1$.}
\label{fig6_Phase_Plot1}
\end{figure}
It is found from Fig. \ref{fig4_ISCO_parameters_co_rotating}, that the ISCO radius $r$ and the corresponding orbital angular momentum $L_{z}$ increase as the parameter $\mathit{C}$ decreases for corotating cases, wehereas for the counter-rotating cases as presented in Fig. \ref{fig5_ISCO_parameters_counter_rotating}, the ISCO parameter $r$ increases, while the ISCO parameter $L_{z}$ decreases with decrease in parameter $\mathit{C}$.
\item Corresponding to the same value of parameter $\mathit{C}$, the ISCO radius for counter-rotating case is more than the corotating case (compare the corresponding part of both the Figs. \ref{fig4_ISCO_parameters_co_rotating} and \ref{fig5_ISCO_parameters_counter_rotating} for reference).
\item The allowed range of the particle spin ${\bf S}$, where the ISCOs exist for the Reissner-Nordstr$\ddot{o}$m like braneworld BH ($a=0$ and $\mathit{C}<0$) increases monotonously with decrease in parameter $\mathit{C}$, whereas for the Reissner-Nordstr$\ddot{o}$m like BH ($a=0$ and $\mathit{C}>0$), the allowed range of the particle spin ${\bf S}$, where the ISCOs exist first decreases and then increases as the parameter $\mathit{C}$ decreases. In Fig. \ref{fig6_Phase_Plot1}, the light yellow and light green regions represented the Reissner-Nordstr$\ddot{o}$m like braneworld BH and the dark yellow and dark green regions represented the Reissner-Nordstr$\ddot{o}$m like BH.
\begin{figure}
\begin{tabular}{c}
\includegraphics[scale=0.72]{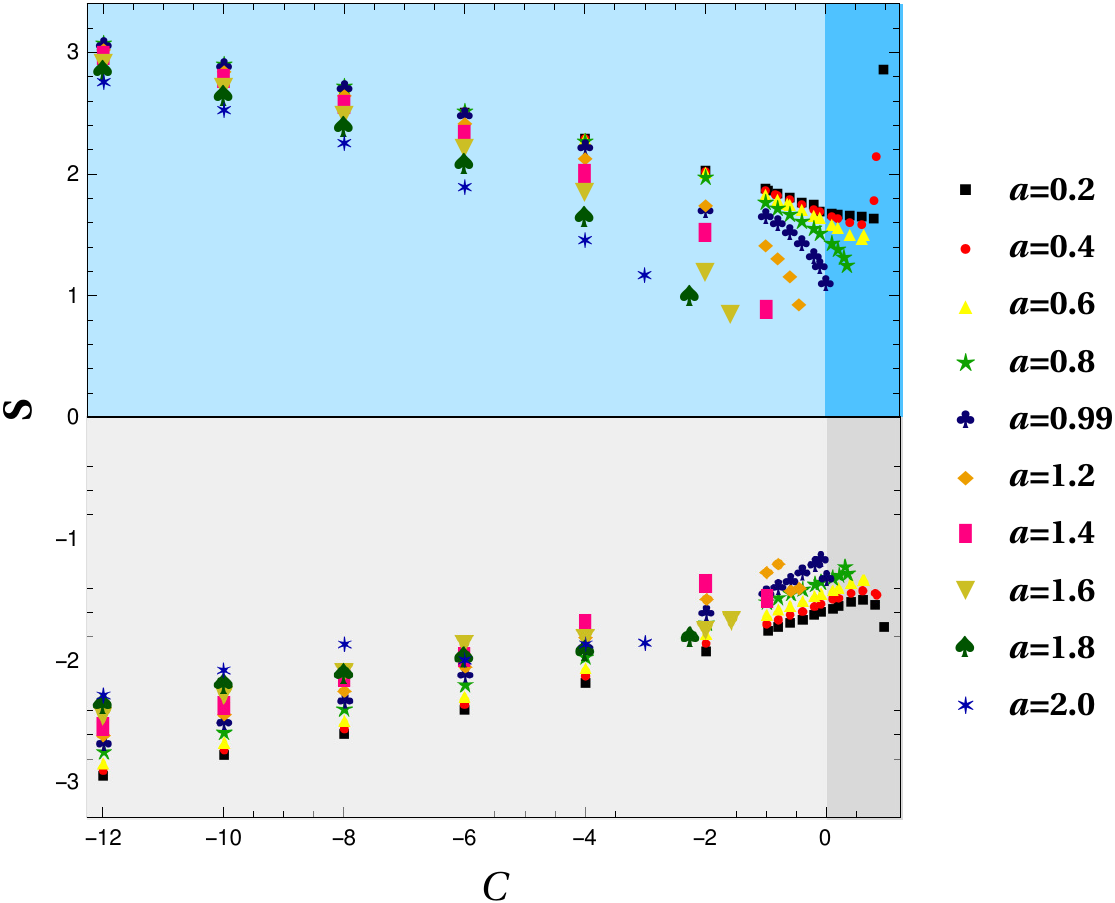}\hspace{0cm}
\end{tabular}
 \caption{(Color-online) Phase plot between the parameters ${\bf S}$ and $C$, shows the bound on the particle's spin ${\bf S}$ for different set of values of parameters $a$ and $C$. The permitted region is between the curves formed by the dots.
 Here, the blue and the grey regions represent corotating ($J_{z}>0$) and counter-rotating ($J_{z}<0$) cases of the spinning particle around rotating braneworld BH, respectively. The mass parameter $M$ sets to unity.}
\label{fig7_Phase_Plot2}
\end{figure}
\item Fig. \ref{fig7_Phase_Plot2}, showed the allowed range of the particle spin ${\bf S}$ for the rotating braneworld BH. The light blue and the light grey regions of rotating braneworld BH correspond to Kerr-Newman like braneworld BH ($a>0$ and $\mathit{C}<0$) whereas dark blue and dark grey regions correspond to Kerr-Newman like BH ($a>0$ and $\mathit{C}>0$). Similar to braneworld Reissner-Nordstr$\ddot{o}$m BH case, the range of the particle spin ${\bf}$ for rotating braneworld BH increases with decrease in parameter $\mathit{C}$. However, for the fixed value of parameter $\mathit{C}$ the allowed range of the parameter ${\bf S}$ decreases as parameter $a$ of BH increases.
\end{itemize}

\section{Summary and conclusion}\label{Conclusion}

In this paper, we have studied in detail the properties of the ergosphere for a rotating braneworld BH and highlight the effect of tidal charge parameter $\mathit{C}$ on it. In Fig. \ref{fig1_ergo}, we showed that its ergoregion (blue) increases with the rise of the tidal charge parameter $\mathit{C}$ for a fixed value of rotation parameter $a$. We also showed that when $\mathit{C}<0$, it is possible to have a rotation parameter greater than unity unlike Kerr and Kerr-Newman BH where the BH rotation parameter $a$ is always less than unity. Hence, for the rotating braneworld BH it is possible to have a super spinning case (ie., $a>1$) as it is also mentioned in \cite{Stuchlik:2017rir,Schee:2008fc,Stuchlik:2008fy}. We also gave the numerical bounds on the rotation parameter $a$ and tidal charge parameter $\mathit{C}$ of a rotating braneworld BH as shown in Fig. \ref{fig2_cont}. It is clear from the figure that the upper bound on the parameter $a$ can be greater than the corresponding Kerr and Kerr-Newman BHs parameter. In Fig. \ref{fig3_veff}, we showed the behavior of effective potential $V_{eff(+)}$ as a function of radial parameter $r$ of the orbit.

Most importantly, we numerically probe the motion of spinning test particle in the rotating braneworld BH background, by solving the system of Eqns. (\ref{appendix_radial_v})-(\ref{appendix_ISCO_cond}) in general ((\ref{radial_v})-(\ref{ISCO_cond})), where we study the  behavior of ISCO parameters for different combinations of values of rotation $a$ and tidal charge $\mathit{C}$ parameters of both corotating and counter-rotating orbit cases around rotating braneworld BH as shown in Figs. \ref{fig4_ISCO_parameters_co_rotating} and \ref{fig5_ISCO_parameters_counter_rotating}. While considering the ISCO parameter $r$ for the rotating braneworld BH, we showed that this ISCO parameter was always greater for counter-rotating orbits than that obtained for corotating orbits for the same value of rotation parameter $a=1.2$.

Our analysis showed some interesting results about the Reissner-Nordstr$\ddot{o}$m and Kerr-Newman BHs, which were not reported in earlier studies \cite{Jefremov:2015gza,Zhang:2017nhl}. Such as $(i)$ it is shown in Fig. \ref{fig6_Phase_Plot1} that the behavior of the particle spin parameter ${\bf S}$ as a function of parameter $\mathit{C}$ is totally symmetric for corotating and counter-rotating orbits around Reissner-Nordstr$\ddot{o}$m like BH (i.e., $Q^{2}=C >0$) under the change $L_z$ to  $- L_z$ and ${\bf S}$ to ${-\bf S}$. $(ii)$ Also, the parameter ${\bf S}$ first decreases and then increases as the parameter $\mathit{C}$ becomes smaller for Reissner-Nordstr$\ddot{o}$m like BH. $(iii)$ Similar to Reissner-Nordstr$\ddot{o}$m like BH case the parameter ${\bf S}$ first decreases and then increases with decrease in parameter $\mathit{C}$, for the small values of rotation parameter $a$ (say when $a \leq 0.6$) as shown in Fig. \ref{fig7_Phase_Plot2}, while the allowed range of parameter ${\bf S}$ always increases with decrease in parameter $\mathit{C}$ when rotation parameter $a \geq 0.8$ for the rotating braneworld BH. Here, it is worth reminding once again that rotating braneworld BH becomes Kerr-Newman like BH when parameter $\mathit{C}>0$.

\section{Acknowledgment}
P.S. would like to thanks Olivier Sarbach for useful discussions at some stage of this work. U. N. acknowledges support from PRODEP-SEP and the CONACYT thematic network project 280908 {\it `Agujeros Negros y Ondas Gravitatorias'} for financial support. R. B. and U. N. acknowledge support from CIC-UMNSH and SNI-CONACYT.
P. S. would like to thank \textit{Programa de Desarrollo Profesional Docente} (PRODEP) of the \textit{Secretar\'{\i}a de Educac\'{\i}on P\'{u}blica} (SEP) of the Mexican government, for providing the financial support through post-doctoral program.

\appendix
\section{THE EXPLICIT FORM OF ISCO EQUATIONS}
\label{appendix}
Here, we present the explicit form of the ISCO Eqs. (\ref{radial_v}) - (\ref{ISCO_cond}) in terms of $x,\; a,\; \mathit{C},\; J_{z}$ and $E$, after doing the transformation $x=1/r$ (for the sake of simplicity):

\begin{strip}
\begin{eqnarray}
\label{appendix_radial_v}
V_{S}(x;a,\mathit{C},E,J_{z})&=&{E}^2 \left(2 a^3 \mathit{C} {\bf S} x^6-2 a^3 {\bf S} x^5-a^2 \mathit{C}^2 {\bf S}^2 x^8+2 a^2 \mathit{C} {\bf S}^2 x^7+2 a^2 \mathit{C} {\bf S}^2 x^6+a^2 \mathit{C} x^4-a^2 {\bf S}^2 x^6\right.\nonumber\\
&&\;\;\;\;\;\;\;\;\;\left.-2 a^2 {\bf S}^2 x^5-2 a^2 x^3-a^2 x^2+4 a \mathit{C} {\bf S} x^4-6 a {\bf S} x^3+\mathit{C} {\bf S}^2 x^4-2 {\bf S}^2 x^3+{\bf S}^2 x^2-1\right)
\nonumber\\
&&+{E} \left(-4 a^2 \mathit{C} {J_{z}} {\bf S} x^6+4 a^2 {J_{z}} {\bf S} x^5+2 a \mathit{C}^2 {J_{z}} {\bf S}^2 x^8-4 a \mathit{C} {J_{z}} {\bf S}^2 x^7-2 a \mathit{C} {J_{z}} {\bf S}^2 x^6-2 a \mathit{C} {J_{z}} x^4\right.
\nonumber\\
&&\;\;\;\;\;\;\;\;\;\left. +2 a {J_{z}} {\bf S}^2 x^6+2 a {J_{z}} {\bf S}^2 x^5+4 a {J_{z}} x^3-4 \mathit{C} {J_{z}} {\bf S} x^4+6 {J_{z}} {\bf S} x^3-2 {J_{z}} {\bf S} x^2\right)
\nonumber\\
&&+a^2 \mathit{C}^2 {\bf S}^4 x^{10}-2 a^2 \mathit{C} {\bf S}^4 x^9+2 a^2 \mathit{C} {\bf S}^2 x^6+a^2 {\bf S}^4 x^8-2 a^2 {\bf S}^2 x^5+a^2 x^2+2 a \mathit{C} {J_{z}}^2 {\bf S} x^6-2 a {J_{z}}^2 {\bf S} x^5
\nonumber\\
&&+\mathit{C}^3 {\bf S}^4 x^{10}-\mathit{C}^2 {J_{z}}^2 {\bf S}^2 x^8-4 \mathit{C}^2 {\bf S}^4 x^9+\mathit{C}^2 {\bf S}^4 x^8+2 \mathit{C}^2 {\bf S}^2 x^6+2 \mathit{C} {J_{z}}^2 {\bf S}^2 x^7+\mathit{C} {J_{z}}^2 x^4
\nonumber\\
&&+5 \mathit{C} {\bf S}^4 x^8-2 \mathit{C} {\bf S}^4 x^7-6 \mathit{C} {\bf S}^2 x^5+2 \mathit{C} {\bf S}^2 x^4+\mathit{C} x^2-{J_{z}}^2 {\bf S}^2 x^6-2 {J_{z}}^2 x^3+{J_{z}}^2 x^2-2 {\bf S}^4 x^7
\nonumber\\
&&+{\bf S}^4 x^6+4 {\bf S}^2 x^4-2 {\bf S}^2 x^3-2 x+1=0,\\
\frac{dV_{S}(x;a,\mathit{C},E,J_{z})}{dx}&=& {E}^2 \left(12 a^3 \mathit{C} {\bf S} x^5-10 a^3 {\bf S} x^4-8 a^2 \mathit{C}^2 {\bf S}^2 x^7+14 a^2 \mathit{C} {\bf S}^2 x^6+12 a^2 \mathit{C} {\bf S}^2 x^5+4 a^2 \mathit{C} x^3-6 a^2 {\bf S}^2 x^5\right.
\nonumber\\
&&\;\;\;\;\;\;\;\;\left. -10 a^2 {\bf S}^2 x^4-6 a^2 x^2-2 a^2 x+16 a \mathit{C} {\bf S} x^3-18 a {\bf S} x^2+4 \mathit{C} {\bf S}^2 x^3-6 {\bf S}^2 x^2+2 {\bf S}^2 x\right)
\nonumber\\
&&+{E} \left(-24 a^2 \mathit{C} {J_{z}} {\bf S} x^5+20 a^2 {J_{z}} {\bf S} x^4+16 a \mathit{C}^2 {J_{z}} {\bf S}^2 x^7-28 a \mathit{C} {J_{z}} {\bf S}^2 x^6-12 a \mathit{C} {J_{z}} {\bf S}^2 x^5\right.
\nonumber\\
&&\;\;\;\;\;\;\;\;\;\left. -8 a \mathit{C} {J_{z}} x^3+12 a {J_{z}} {\bf S}^2 x^5+10 a {J_{z}} {\bf S}^2 x^4+12 a {J_{z}} x^2-16 \mathit{C} {J_{z}} {\bf S} x^3+18 {J_{z}} {\bf S} x^2-4 {J_{z}} {\bf S} x\right)
\nonumber\\
&&+10 a^2 \mathit{C}^2 {\bf S}^4 x^9-18 a^2 \mathit{C} {\bf S}^4 x^8+12 a^2 \mathit{C} {\bf S}^2 x^5+8 a^2 {\bf S}^4 x^7-10 a^2 {\bf S}^2 x^4+2 a^2 x+12 a \mathit{C} {J_{z}}^2 {\bf S} x^5
\nonumber\\
&&-10 a {J_{z}}^2 {\bf S} x^4+10 \mathit{C}^3 {\bf S}^4 x^9-8 \mathit{C}^2 {J_{z}}^2 {\bf S}^2 x^7-36 \mathit{C}^2 {\bf S}^4 x^8+8 \mathit{C}^2 {\bf S}^4 x^7+12 \mathit{C}^2 {\bf S}^2 x^5
\nonumber\\
&&+14 \mathit{C} {J_{z}}^2 {\bf S}^2 x^6+4 \mathit{C} {J_{z}}^2 x^3+40 \mathit{C} {\bf S}^4 x^7-14 \mathit{C} {\bf S}^4 x^6-30 \mathit{C} {\bf S}^2 x^4+8 \mathit{C} {\bf S}^2 x^3+2 \mathit{C} x
\nonumber\\
&&-6 {J_{z}}^2 {\bf S}^2 x^5-6 {J_{z}}^2 x^2+2 {J_{z}}^2 x-14 {\bf S}^4 x^6+6 {\bf S}^4 x^5+16 {\bf S}^2 x^3-6 {\bf S}^2 x^2-2=0,\\
\label{appendix_radial_a}
\frac{d^{2}V_{S}(x;a,\mathit{C},E,J_{z})}{dx^{2}}&=& {E}^2 \left(60 a^3 \mathit{C} {\bf S} x^4-40 a^3 {\bf S} x^3-56 a^2 \mathit{C}^2 {\bf S}^2 x^6+84 a^2 \mathit{C} {\bf S}^2 x^5+60 a^2 \mathit{C} {\bf S}^2 x^4+12 a^2 \mathit{C} x^2-30 a^2 {\bf S}^2 x^4\right.
\nonumber\\
&&\left. -40 a^2 {\bf S}^2 x^3-12 a^2 x-2 a^2+48 a \mathit{C} {\bf S} x^2-36 a {\bf S} x+12 \mathit{C} {\bf S}^2 x^2-12 {\bf S}^2 x+2 {\bf S}^2\right)
\nonumber\\
&&+{E} \left(-120 a^2 \mathit{C} {J_{z}} {\bf S} x^4+80 a^2 {J_{z}} {\bf S} x^3+112 a \mathit{C}^2 {J_{z}} {\bf S}^2 x^6-168 a \mathit{C} {J_{z}} {\bf S}^2 x^5-60 a \mathit{C} {J_{z}} {\bf S}^2 u^4\right.
\nonumber\\
&&\left. -24 a \mathit{C} {J_{z}} x^2+60 a {J_{z}} {\bf S}^2 x^4+40 a {J_{z}} {\bf S}^2 x^3+24 a {J_{z}} x-48 \mathit{C} {J_{z}} {\bf S} x^2+36 {J_{z}} {\bf S} x-4 {J_{z}} {\bf S}\right)
\nonumber\\
&&+90 a^2 \mathit{C}^2 {\bf S}^4 x^8-144 a^2 \mathit{C} {\bf S}^4 x^7+60 a^2 \mathit{C} {\bf S}^2 x^4+56 a^2 {\bf S}^4 x^6-40 a^2 {\bf S}^2 x^3+2 a^2+60 a \mathit{C} {J_{z}}^2 {\bf S} x^4
\nonumber\\
&&-40 a {J_{z}}^2 {\bf S} x^3+90 \mathit{C}^3 {\bf S}^4 x^8-56 \mathit{C}^2 {J_{z}}^2 {\bf S}^2 x^6-288 \mathit{C}^2 {\bf S}^4 x^7+56 \mathit{C}^2 {\bf S}^4 x^6+60 \mathit{C}^2 {\bf S}^2 x^4
\nonumber\\
&&+84 \mathit{C} {J_{z}}^2 {\bf S}^2 x^5+12 \mathit{C} {J_{z}}^2 x^2+280 \mathit{C} {\bf S}^4 x^6-84 \mathit{C} {\bf S}^4 x^5-120 \mathit{C} {\bf S}^2 x^3+24 \mathit{C} {\bf S}^2 x^2+2 \mathit{C}
\nonumber\\
&&-30 {J_{z}}^2 {\bf S}^2 x^4-12 {J_{z}}^2 x+2 {J_{z}}^2-84 {\bf S}^4 x^5+30 {\bf S}^4 x^4+48 {\bf S}^2 x^2-12 {\bf S}^2 x=0.
\label{appendix_ISCO_cond}
\end{eqnarray}
\end{strip}
It is worth mentioning here that working with parameter $x$ instead of parameter $r$ does not change the form of Eqs. (\ref{radial_v}) - (\ref{ISCO_cond}) as shown in \cite{Jefremov:2015gza} and hence the system comprising of Eqs. (\ref{radial_v}) - (\ref{ISCO_cond}) is identical to the system of Eqs. (\ref{appendix_radial_v}) - (\ref{appendix_ISCO_cond}).


\end{document}